\documentclass[aps,prl,graphicx,9pt,twocolumn,superscriptaddress]{revtex4-2}
\usepackage{amsmath,amscd,graphicx,graphicx,cases}
\usepackage{multirow,float,subfigure,appendix}
\usepackage{diagbox,makecell,blindtext,color,amssymb}
\usepackage[colorlinks,linkcolor=blue,urlcolor=blue,
anchorcolor=blue,citecolor=blue]{hyperref}
\usepackage{textcomp,booktabs,colortbl,easyReview}
\newcolumntype{C}{c<{\kern\tabcolsep}@{}}
\usepackage{times}
\usepackage{lineno}
\usepackage{xcolor,ulem,bm}
\usepackage{orcidlink}

\begin{document}

\title{Experimental demonstration of spontaneous symmetry breaking with emergent multi-qubit entanglement}

\author{Ri-Hua Zheng}
\thanks{These authors contribute equally to this work.}
\author{ Wen Ning}
\thanks{These authors contribute equally to this work.}\author{Jia-Hao L\"{u}} 
\author{Xue-Jia Yu}
\thanks{These authors contribute equally to this work.}

\author{Fan Wu}
\affiliation{Fujian Key Laboratory of Quantum Information and Quantum Optics, College of Physics and Information Engineering, Fuzhou University, Fuzhou, Fujian, 350108, China}

\author{ Cheng-Lin Deng }
\affiliation{Institute of Physics and Beijing National Laboratory for Condensed Matter Physics, Chinese Academy of Sciences, Beijing 100190, China}
\affiliation{CAS Center for Excellence in Topological Quantum Computation, University of Chinese Academy of Sciences, Beijing 100190, China}

\author{ Zhen-Biao Yang}\thanks{E-mail: zbyang@fzu.edu.cn}
\affiliation{Fujian Key Laboratory of Quantum Information and Quantum Optics, College of Physics and Information Engineering, Fuzhou University, Fuzhou, Fujian, 350108, China}
\affiliation{Hefei National Laboratory, Hefei 230088, China}

\author{Kai Xu}\thanks{E-mail: kaixu@iphy.ac.cn}
\author{ Dongning Zheng}
\author{ Heng Fan}
\affiliation{Institute of Physics and Beijing National Laboratory for Condensed Matter Physics, Chinese Academy of Sciences, Beijing 100190, China}
\affiliation{CAS Center for Excellence in Topological Quantum Computation, University of Chinese Academy of Sciences, Beijing 100190, China}
\affiliation{Hefei National Laboratory, Hefei 230088, China}

\author{Shi-Biao Zheng}\thanks{E-mail: t96034@fzu.edu.cn}
\affiliation{Fujian Key Laboratory of Quantum Information and Quantum Optics, College of Physics and Information Engineering, Fuzhou University, Fuzhou, Fujian, 350108, China}
\affiliation{Hefei National Laboratory, Hefei 230088, China}
\thanks{E-mail: zbyang@fzu.edu.cn}

\begin{abstract}

Spontaneous symmetry breaking (SSB) is crucial to the occurrence of phase transitions. 
Once a phase transition occurs, a quantum system presents degenerate eigenstates that lack the symmetry of the Hamiltonian. 
After crossing the critical point, the system is essentially evolved to a quantum superposition of these eigenstates until decoherence sets in. 
Despite the fundamental importance and potential applications in quantum technologies, such quantum-mechanical SSB phenomena have not been experimentally explored in many-body systems. 
We here present an experimental demonstration of the SSB process in the Lipkin-Meshkov-Glick model, governed by the competition between the individual driving and intra-qubit interaction. 
The model is realized in a circuit quantum electrodynamics system, where 6 Xmon qubits are coupled in an all-to-all manner through virtual photon exchange mediated by a resonator. 
The observed nonclassical correlations among these qubits in the symmetry-breaking region go beyond the conventional description of SSB, shedding new light on phase transitions for quantum many-body systems.

\end{abstract}

\maketitle

\narrowtext

\bigskip The idea of spontaneous symmetry breaking (SSB) lies at the heart of the Landau-Ginzburg-Wilson paradigm for describing phase transitions \cite{Sachdev2011}. 
This idea was first proposed in solid physics, and is crucial to the understanding of phenomena in condensed matter physics, e.g., superconductivity \cite{Bardeen1957}. 
In the 1960s, the SSB mechanism of the BCS theory of superconductivity was transported to particle physics \cite{Nambu1960, Baker1962, Higgs1964, Englert1964, Guralnik1964}, successfully explaining the origin of the mass of non-Abelian gauge bosons carrying weak forces. 
This breakthrough rendered the Yang-Mills gauge theory \cite{Yang1954} to become a powerful framework for the foundation of the Standard Model \cite{Weinberg2004}, where the electromagnetic, weak, and strong interactions are unified. SSB also plays an important role in the evolution of the universe. 
During the expansion and coolingdown of the universe, it might have undergone SSB phase transitions, which transformed the ``vacuum" from a state with higher symmetries to one with lower symmetries \cite{Kibble1980, Nambu2009}.

The most typical example for illustrating SSB is a classical model, where a point-like particle moves along a straight line under the influence of a sombrero potential, given by $V(x)=\Lambda x^{4}-\mu x^{2}$ with $\Lambda,\mu >0$ \cite{Munoz-Vega2012}. 
The particle has two different positions of stable equilibrium, $x=\pm \sqrt{\mu /2\Lambda }$. Although the potential possesses a global spatial-inversion symmetry, i.e., $V(-x)=V(x)$, each of the two equilibrium positions loses this symmetry. 
Which equilibrium position the particle chooses to stay at depends on perturbation. 
This is the standard understanding of SSB, which, however, does not grasp the quantum-mechanical feature of SSB. 
For quantum systems, the superposition principle allows degenerate stable symmetry-breaking eigenstates to be superimposed with each other, forming a symmetric one. As such, SSB is said to occur if there exist two or more degenerate eigenstates, each of which lacks the symmetry of the governing Hamiltonian \cite{Munoz-Vega2012}. 
When the system initially has the same symmetry as the Hamiltonian, it should evolve to the symmetric superposition of the corresponding degenerate eigenstates after a quantum phase transition (QPT) under the Hamiltonian dynamics. 
For an open quantum system, it is the environment-induced decoherence that destroys the quantum coherence, forcing the system to choose one of the symmetry-breaking eigenstates as the reality \cite{Zurek1991, Zurek2003, Dziarmaga2012}. 

During the past few years, QPT have been probed in different many-body systems, including neutral atom arrays \cite{Lukin2019, Ebadi2021}, trapped ions \cite{Jurcevic2017, Zhang2017, Li2022, Joshi2023}, superconducting qubits \cite{Xu2018, Xu2020, Gong2021}, and ultracold atoms trapped in optical cavities \cite{Baumann2010, Baumann2011, Leonard2017, Leonard2017_2, Klinder2015, Zhang2017_2, Zhang2021}. 
In particular, symmetry-breaking phases have been experimentally explored in Dicke-like QPTS \cite{Baumann2011, Leonard2017, Leonard2017_2}.
However, the driven-dissipative character of the realized phase transitions makes it impossible to observe the quantum coherence between the SSB degenerate eigenstates. 
So far, experimental demonstrations of SSB with quantum-mechanical features have been confined to a photonic mode that is coupled to a single qubit \cite{Zheng2023, Ning2024}. 
For a finite many-qubit system, one of the most essential characteristics of the SSB phase transition is the appearance of a multi-qubit maximally entangled state \cite{Zeng2015, Zeng2019, Verstraete2004}, also referred to as the Greenberger-Horne-Zeilinger (GHZ) state \cite{Greenberger1989}. 
{Although quantum entanglement has been produced by adiabatic control of the quantum Ising Hamiltonian in a two-qubit ion trap experiment \cite{Friedenauer2008}, such a highly nonclassical SSB feature has not been experimentally observed in QPTs of many-body systems.}

{We here demonstrate the purely quantum-mechanical SSB, featuring emergent GHZ entanglement, in the QPT of the multi-qubit Lipkin-Meshkov-Glick (LMG) model \cite{Lipkin1965, Latorre2005, Huang2018, Louren2020, Bao2020}, governed by a Hamiltonian featuring the competition between individual continuous drivings and intra-qubit interactions.}
The LMG model is synthesized in a superconducting circuit, where 6 Xmon qubits are nearly homogeneously coupled to one another with the assistance of a bus resonator connected to all these qubits. 
By slowly decreasing the driving strength, we realize the SSB process, during which the multi-qubit system evolves from an initial symmetric product state to a superposition of two degenerate symmetry-breaking eigenstates of the governing Hamiltonian. 
Through the real-time probing of the average two-qubit longitudinal correlation during the {quasi-adiabatic} process, we observe the progressive improvement of the $Z_{2}$ symmetry breaking degree.
We verify the quantum character of the SSB by measuring multi-qubit transverse quantum correlations and mapping out the Wigner function. 
{Our results reveal the fundamental distinction between the SSB in QPT of many-body systems and that in classical phase transitions}, and {offer a dephasing-insensitive strategy for multi-qubit entanglement engineering in solid-state systems, thanks to the dynamical decoupling effects produced by drivings individually applied to the qubits \cite{Guo2018}.}

The system under consideration corresponds to the $N$-qubit LMG model, featuring the competition of the individual drivings and qubit-qubit swapping couplings. 
The system dynamics is governed by the Hamiltonian (setting $\hbar =1$)
\begin{eqnarray}\label{Heff}
H=-\Omega S_{x}-\lambda S_{z}^{2}, 
\end{eqnarray}
where $\Omega $ is the driving strength, $\lambda $ denotes the qubit-qubit coupling, $S_{x/z}=\frac{1}{2}\stackrel{N}{\mathrel{\mathop{\sum }\limits_{j=1}}}\sigma _{j}^{x/z}$ is the collective angular operator, and $\sigma_{j}^{x/z}$ denotes the Pauli operators of the $j$th qubit. 
The Hamiltonian has the $Z_{2}$ symmetry, defined as the joint parity: ${\cal P}_{x}=\otimes_{j=1}^{N}\sigma _{j}^{x}$. 
The system's ground-state behavior depends on the competition of longitudinal couplings and transverse drivings. 
When $\Omega, \lambda>0$ and $\eta=N\lambda/\Omega \ll 1$, the longitudinal couplings are suppressed, so that the system presents a unique ground state of the form $\left\vert X\right\rangle =\otimes_{j=1}^{N}\left\vert +_{j}\right\rangle $, where $\left\vert \pm _{j}\right\rangle =\frac{1}{\sqrt{2}}(\left\vert g_{j}\right\rangle \pm \left\vert e_{j}\right\rangle )$. 
Here $\left\vert g_{j}\right\rangle $ and $\left\vert e_{j}\right\rangle $ denote the lower and upper levels of the $j$th qubit. 
This eigenstate has the same symmetry as the Hamiltonian, i.e., ${\cal P}_{x}\left\vert X\right\rangle =\left\vert X\right\rangle $. 
When $\eta \gg 1 $, the transverse driving only slightly shifts the energy levels of the eigenstates $\left\vert J,m\right\rangle $ ($\left\vert J,m\right\rangle $ is the eigenstate of $S_z$ with eigenvalue $m$ and total angular momentum quantum number $J$. See Sec. S1 of the Supplementary Material \cite{supp} for the detailed derivation).
The most remarkable feature of the system in the strong-coupling regime is that it has two degenerate ground states, $\left\vert J,J\right\rangle=\otimes _{j=1}^{N}\left\vert e_{j}\right\rangle $ and $\left\vert J,-J\right\rangle =\otimes _{j=1}^{N}\left\vert g_{j}\right\rangle $, each breaking the system's $Z_{2}$ symmetry. 

Suppose that the system is initially in the state $\left\vert X\right\rangle$, which is the lowest eigenstate of the Hamiltonian under the condition $\eta \ll 1$. When the control parameter $\Omega $ is slowly decreased, the SSB will occur. 
In the limit $\eta \gg 1$, the system will be moved to the symmetric superposition of the two degenerate ground states, $\left\vert \psi _{+}\right\rangle =\frac{1}{\sqrt{2}}(\left\vert J,J\right\rangle +\left\vert J,-J\right\rangle )$, which corresponds to an $N$-qubit GHZ state. 
As the slow variation of the transverse drive does not change the symmetry of the Hamiltonian, the simultaneous emergence of the two symmetry-breaking eigenstates is a result of SSB QPT. 
{When $N$ is sufficiently large, the QPT occurs at the critical point $\eta=1$. For the six-qubit case, the system's state is not suddenly changed at this point, but the GHZ entanglement still emerges when $\eta\gg 1$ as a consequence of SSB (see Sec. S2 of the Supplemental Material \cite{supp} for details).}
The SSB is manifested by the vulnerable entanglement properties of the GHZ state, which will be transformed into a classical mixture by the environment-induced decoherence within a timescale $\tau_{\rm dec}\sim 1/N$ \cite{Dziarmaga2012}. 
We note that the adiabatic preparation of superpositions of two degenerate ground states has been demonstrated in Rydberg atom arrays \cite{Omran2019}, but where the emergence of the degeneracy cannot be attributed to SSB as the Hamiltonian itself has no symmetry.

The LMG model can be engineered with a circuit quantum electrodynamics (QED) architecture involving $N$ qubits with the same frequency $\omega$, each interacting with a microwave resonator with a fixed frequency $\omega_{b}$, and subjected to a continuous drive with the frequency $\omega_{o}$, the strength $\Omega $, and the phase $\pi $. 
Under the condition that $\omega-\omega _{b}$ is much larger than the qubit-resonator coupling strength $\xi$, the resonator will not exchange photons with the qubits, but can induce couplings between any pair of qubits through virtual photon exchange \cite{Zheng2000, Agarwal1997, Zheng2001, Song2017, Song2019}. 
When the resonator is initially in the vacuum state, it will remain in this state throughout the interaction, so that it can be discarded in the description of the state evolution of the qubits. 
Discarding a trivial constant term, the resonator-induced qubit-qubit couplings are described by the effective Hamiltonian $ \lambda  (-S_{z}^{2}+S_{z})$, where $\lambda =\xi^{2}/|\omega_{o}-\omega_{b}|$. 
When the frequency of the continuous drive satisfies $\omega _{o}=\omega + \lambda$, the system Hamiltonian in the framework rotating at the frequency $\omega_{o}$ is described by Eq. (\ref{Heff}).

\begin{figure}[t]
  \centering
  \includegraphics[width=8.5cm]{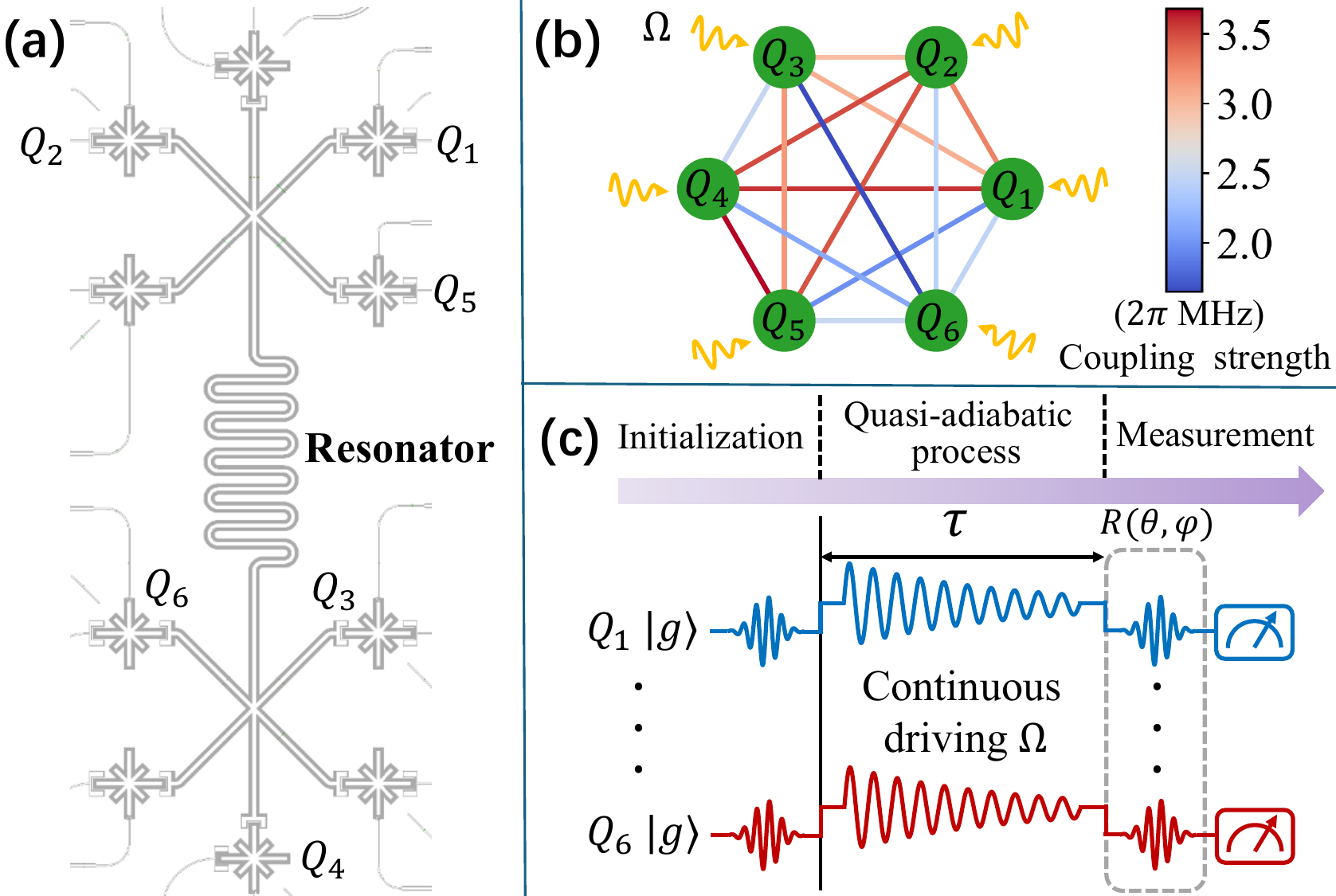} 
  \caption{Experimental realization of SSB in a multi-qubit system. (a) Device schematic. The device comprises 10 Josephson-junction-based qubits and a bus resonator. 
  The first 6 qubits, denoted as $Q_{j}$ ($j=1$ to $6$), are approximately homogeneously coupled to the resonator with the on-resonance photonic swapping rate $\xi\simeq 2\pi \times 20$ MHz.
  The other four qubits are unused in the experiment. (b) Synthesis of the LMG model. When these qubits are detuned from the resonator by the same amount $|\Delta| \gg \xi $, any pair of qubits,  $Q_{j}$ and $Q_{k}$, are coupled through virtual photon exchange mediated by the resonator, with the effective coupling strength $\lambda \simeq \xi^{2}/|\Delta| $. 
  Each of these qubits is driven by a continuous microwave with the Rabi frequency $\Omega$. 
  The system ground state depends upon the competition between resonator-induced qubit-qubit couplings and transverse drivings. 
  (c) Pulse sequence. 
  The experiment starts by tuning the qubits to the corresponding idle frequencies, where they are prepared in the state $\otimes _{j=1}^{6}\left\vert +_{j}\right\rangle $ with a continuous microwave pulse. 
  By tuning all the qubits to the same operation frequency and applying continuous drivings, the system dynamics is described by the LMG Hamiltonian. 
  Following a preset {quasi-adiabatic} process, the qubits are biased back to their idle frequencies for the state readout.
  } 
  \label{fig1}
\end{figure}

\bigskip Our circuit QED system is composed of 10 frequency-tunable superconducting qubits ($Q_{j}$), coupled to a bus resonator with a fixed frequency $\omega_{b}/(2\pi)\simeq 5.796$ GHz. 
The LMG model is realized with the first 6 qubits that are almost homogeneously coupled to the resonator with the photonic swapping rate $\xi \simeq 2\pi \times20$ MHz, as sketched in Fig. \ref{fig1}\textcolor{blue}{(a)}.
The other 4 unused qubits (not shown) are highly detuned from the resonator and the first 6 qubits, so that they do not affect the engineered LMG model. 
The energy decaying times ($T_{1}$) of these qubits range from 9 to 32 $\mu$s, while the dephasing times ($T_{2}$) range from 2 to 5 $\mu$s. 
The system parameters are detailed in Sec. S3 of the Supplementary Material \cite{supp} (see also Ref. \cite{Song2017}). 
As the qubits' maximum frequencies in our system are close to $\omega _{b}$, the LMG model is synthesized at the operation frequency $\omega _{o}/(2\pi)=5.6895$ GHz, which is lower than $\omega _{b}/(2\pi)$ by an amount of 106.5 MHz [Fig. \ref{fig1}\textcolor{blue}{(b)}]. The resulting intra-qubit couplings induced by the virtual photon exchange have a magnitude of $\lambda=2\pi \times 3.8$ MHz. 
The combination of these effective swapping couplings and transverse drivings with a zero phase is described by the effective Hamiltonian $-H$. 
The SSB QPT associated with the highest eigenstate of $-H$ is equivalent to the ground phase transition of $H$. 
Before the experiment, all qubits are initialized to their ground states at the idle frequencies. The pulse sequence for realizing the QPT is shown in Fig. \ref{fig1}\textcolor{blue}{(c)}. 
The experiment starts by tuning all these qubits to their idle frequencies.
Then a $\pi /2$ pulse is applied to each qubit, transforming it to the superposition state $\left\vert +_{j}\right\rangle =\frac{1}{\sqrt{2}}(\left\vert g_{j}\right\rangle +\left\vert e_{j}\right\rangle )$. 
This is followed by tuning all the qubits to the operation frequency and applying continuous drivings, incorporating the transverse drivings with the longitudinal couplings.
The phase transition is realized by a {quasi-adiabatic} process shown in Fig. \ref{fig2}\textcolor{blue}{(a)}. 
{As shown in Sec. S5 of the Supplemental Material \cite{supp}, the adiabatic condition can be well satisfied during this process.}
Then each qubit is biased back to its idle frequency for the state readout. 

\begin{figure}[t]
  \centering
  \includegraphics[width=8.5cm]{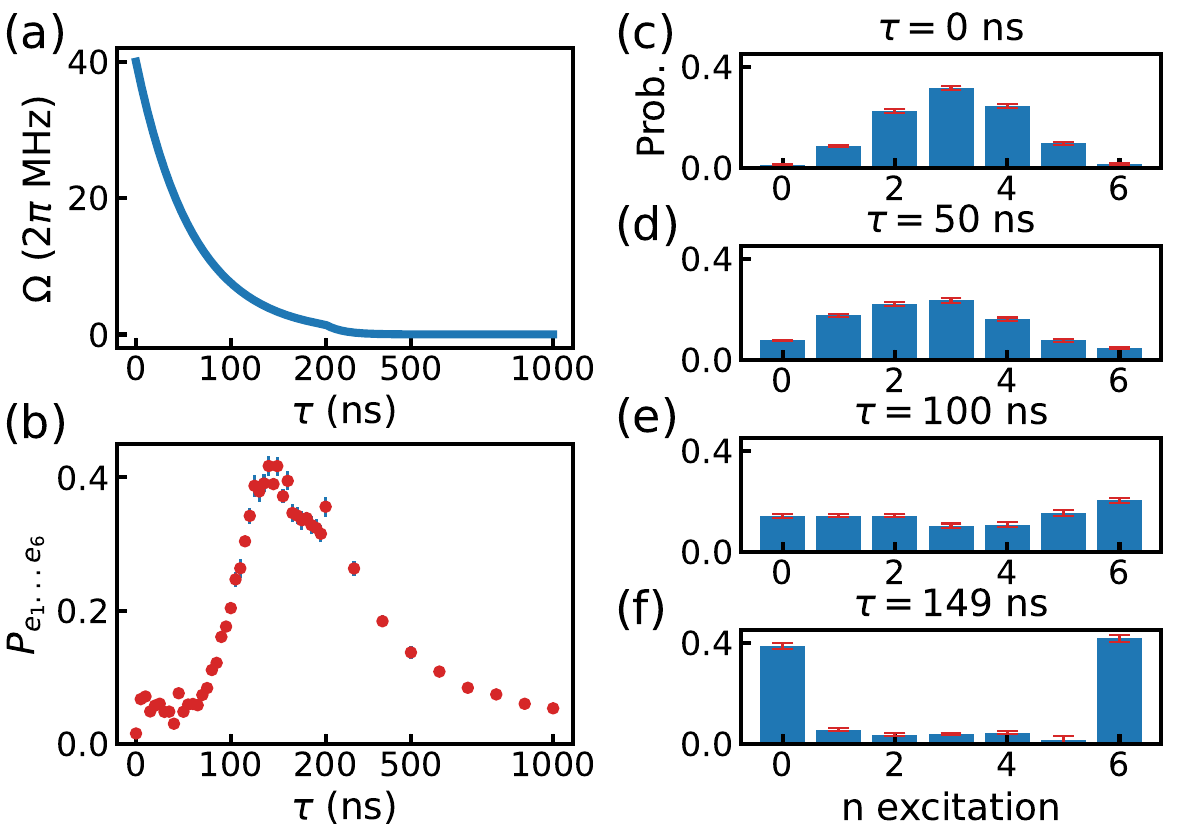} 
  \caption{Observed populations evolutions. (a) $\Omega $ versus $\tau $ during the {quasi-adiabatic} process, varied  as $\Omega (\tau )=\Omega^0 \exp(-\tau/t_f)$,  where $\Omega^0/(2\pi)=40$ MHz, $t_f=60$ ns, and $\tau $ represents the time.
  (b) Measured population, $P_{e_1...e_6}$, as a function of the time. (c), (d), (e), (f) Population distributions measured at $\tau =0$ ns (c), $\tau =50$ ns (d), $\tau =100$ ns (e), and $\tau =149$ ns (f). 
  For clarity, we combine all populations with the same excitation number $n$ into a single one.} 
  \label{fig2}
\end{figure}

To characterize the state evolution, it is necessary to be able to distinguish between $\left\vert \psi _{0}\right\rangle =\left\vert +_{j}\right\rangle ^{\otimes 6}$ and $\left\vert \psi _{1}\right\rangle =\frac{1}{\sqrt{2}}(\left\vert g_{j}\right\rangle ^{\otimes 6}+\left\vert e_{j}\right\rangle ^{\otimes 6})$ that are highest eigenstates in the limits $\eta \ll 1$ and $\eta \gg 1$, respectively. 
Though both states possess a zero z-axis magnetization $\sum_{j=1}^{N}\left\langle \sigma _{j}^{z}\right\rangle $, they have very different population distributions. For $\left\vert \psi _{0}\right\rangle $, the probability of all qubits being populated in $\left\vert e\right\rangle $ is $P_{e_1...e_6}=1/2^{6}$, while that for $\left\vert \psi_{1}\right\rangle $ is $1/2$. 
The measured population, $P_{e_1...e_6}$, as a function of the time, is displayed in Fig. \ref{fig2}\textcolor{blue}{(b)}. 
{As expected, $P_{e_1...e_6}$ is increased monotonously with the increase of $\eta$ before $\tau=149$ ns. The progressive decrease of this population after $149$ ns is due to the energy relaxations of the qubits.} 
To reveal the system evolution more detailedly, we display the population distributions measured at $\tau =0$, $\tau =50$, $\tau =100$, and $\tau =149$ ns in Figs. \ref{fig2}\textcolor{blue}{(c)}, \ref{fig2}\textcolor{blue}{(d)}, \ref{fig2}\textcolor{blue}{(e)} and \ref{fig2}\textcolor{blue}{(f)}, respectively, which confirm the system population undergoes a significant change during the {quasi-adiabatic process}, transformed from a binomial distribution to a pattern dominated by the two maximally distinct components $\left\vert g_{j}\right\rangle^{\otimes 6}$ and $\left\vert e_{j}\right\rangle ^{\otimes 6}$. 
Each of these components corresponds to a symmetry-breaking order with respect to the z-axis magnetization. Due to the environment-induced decays, $\left\vert g_{j}\right\rangle^{\otimes 6}$ is chosen as the classical reality after a long time dynamics. 

\begin{figure}[t]
  \centering
  \includegraphics[width=8.5cm]{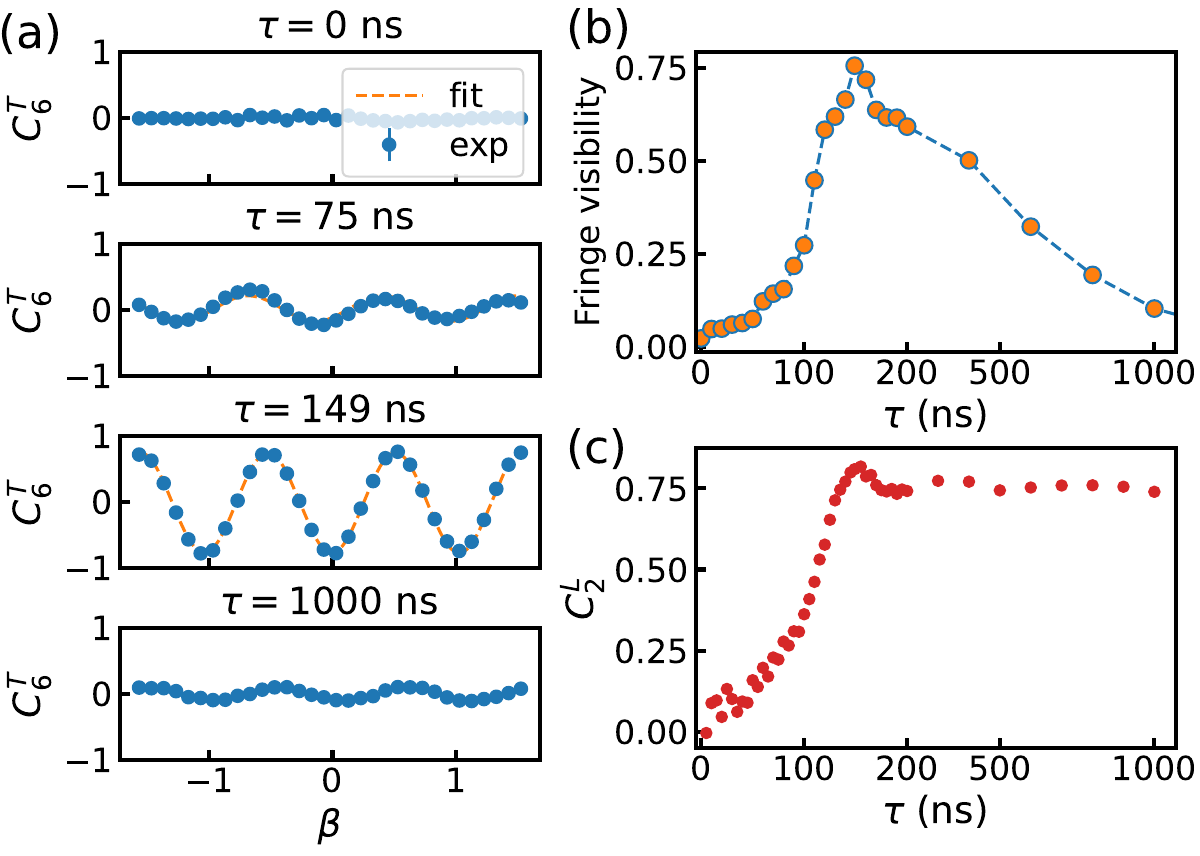} 
  \caption{Measured quantum correlations. 
  (a) 6-qubit transverse quantum correlation ${\cal C}_{6}^{T}$ versus $\beta $, measured for  times $\tau =0$, 75, 149, and 1000 ns. 
  Here ${\cal C}_{6}^{T}$ is defined as $\left\langle \otimes _{j=1}^{6}\sigma _{j}^{\beta }\right\rangle-\left\langle \sigma _{j}^{\beta }\right\rangle ^{\otimes 6}$, where $\sigma_{j}^{\beta }$ is the spin component along the axis with an angle $\beta$ to the x-axis on the equatorial plane of the Bloch sphere. 
  (b) Fringe visibility of the measured ${\cal C}_{6}^{T}$ as a function of $\tau $. 
  (c) Average 2-qubit longitudinal correlation ${\cal C}_{2}^{L}$ versus $\tau $.} 
  \label{fig3}
\end{figure}

These populations do not reflect the quantum coherence between $\left\vert g_{j}\right\rangle ^{\otimes 6}$ and $\left\vert e_{j}\right\rangle^{\otimes 6}$. 
This coherence can be characterized by the nonvanishing quantum correlations among the transverse components of the Bloch vectors, denoted as 
\begin{eqnarray}
{\cal C}_{6}^{T}=\left\langle (\sigma _{j}^{\beta })^{\otimes
6}\right\rangle -\left\langle \sigma _{j}^{\beta }\right\rangle ^{\otimes
6}, 
\end{eqnarray}
where $\sigma _{j}^{\beta }=\cos \beta \sigma _{j}^{x}+\sin \beta \sigma_{j}^{y}$. 
To measure ${\cal C}_{6}^{T}$, before the state readout, each qubit is subjected to a $\pi /2$ rotation around the axis with an angle $\beta -\pi/2 $ to the x-direction on the equatorial plane. The off-diagonal element of the system density matrix, $\rho _{e,...,e;g,...,g}$, is directly related to  ${\cal C}_{6}^{T}$, by ${\cal C}_{6}^{T}=2\left\vert \rho _{e,...,e;g,...,g}\right\vert \cos (6\beta -\gamma )$, where $\gamma =\arg \rho _{e,...,e;g,...,g}$ \cite{Song2017, Monz2011}. 
Figure \ref{fig3}\textcolor{blue}{(a)} shows ${\cal C}_{6}^{T}$ versus $\beta $, measured for different times, from which the visibility of the measured ${\cal C}_6^{T}$ as a function of $\tau $ is inferred and shown in Fig. \ref{fig3}\textcolor{blue}{(b)}. 
The results show that the quantum coherence $\left\vert \rho_{e,...,e;g,...,g}\right\vert $ does not show a monotonous behavior during the {quasi-adiabatic process}, which can be interpreted as a consequence of the competition between the coherent evolution and the relaxation. 
The Hamiltonian dynamics renders the system population to converge to $\left\vert g_{j}\right\rangle^{\otimes N}$ and $\left\vert e_{j}\right\rangle ^{\otimes N}$, while the relaxation deteriorates their quantum coherence. 
At the beginning, the populations of $\left\vert g_{j}\right\rangle ^{\otimes 6}$ and $\left\vert e_{j}\right\rangle^{\otimes 6}$ are almost $0$, and so is their coherence.
As the amount of coherence lost during a given time is proportional to the coherence itself, at the early stage the Hamiltonian dynamics dominates the relaxation. With the increase of the populations of $\left\vert g_{j}\right\rangle ^{\otimes 6}$ and $\left\vert e_{j}\right\rangle^{\otimes 6}$, the relaxation plays a more and more important role. 
When the decoherence-induced coherence loss exceeds the gain due to the Hamiltonian dynamics, ${\cal C}_{6}^{T}$ begins to descend{, and tends to 0 after a long time evolution as a consequence of decoherence}. 
The maximum fringe contrast of ${\cal C}_{6}^{T}$ is 0.75, which, together with the measured populations of $\left\vert g_{j}\right\rangle ^{\otimes 6}$ (0.38) and $\left\vert e_{j}\right\rangle ^{\otimes 6}$ (0.41), yields a fidelity of 0.77 with respect to the ideal 6-qubit GHZ state. 
{The sources of experimental errors include the nonadiabatic effect, qubits' dephasings and energy relaxations, inhomogeneity of the intra-qubit couplings, and imperfections in qubit control. The errors caused by these imperfections are individually quantified in Sec. S6 of the Supplemental Material \cite{supp}.}

{The emergent GHZ state can exhibit multipartite and high-order quantum correlations beyond two-qubit entanglement \cite{Greenberger1989, Louren2020, Girolami2017, Bao2020}. Furthermore, such a state only has a 1/32 overlapping with the initial state, which reflects a drastic change of the system state during the adiabatic evolution. In distinct contrast, for the two-qubit case the overlapping between the initial and final state is as high as 0.5 \cite{Friedenauer2008}. Recently, multi-qubit ground states of Ising models were quasi-adiabatically prepared with trapped-ion simulators \cite{Guo2024, Qiao2024}, but where multipartite quantum correlations, which represent the most remarkable nonclassical feature of SSB quantum phase transitions, were not measured.}

The transverse quantum correlation can measure the extent to which the symmetry is broken only when there are no decoherence effects, as it is zero for any classical mixture of the two symmetry-breaking components $\left\vert g_{j}\right\rangle ^{\otimes 6}$ and $\left\vert e_{j}\right\rangle ^{\otimes 6}$. 
To quantify the symmetry-breaking extent for a general multi-qubit state, we use the average two-qubit longitudinal correlation, defined as 
\begin{eqnarray}
{\cal C}_{2}^{L}=\frac{1}{15}\sum_{k>j}\sum_{j=1}^{6}\left\langle \sigma_{j}^{z}\sigma _{k}^{z}\right\rangle . 
\end{eqnarray}
This longitudinal correlation is equal to 0 for the $Z_{2}$-symmetric eigenstate $|X\rangle$, and equals to 1 for any combination of the $Z_{2}$-symmetry-breaking eigenstates $\left\vert g_{j}\right\rangle ^{\otimes 6}$ and $\left\vert e_{j}\right\rangle ^{\otimes 6}$, independent of their quantum coherence. 
This implies that ${\cal C}_{2}^{L}$ can be used as a measure to characterize the SSB degree. 
For the ideal LMG model, ${\cal C}_{2}^{L}$ would monotonously increase when $\Omega$ is adiabatically decreased, tending to 1 when $\eta \to \infty$. 
Figure \ref{fig3}\textcolor{blue}{(c)} displays ${\cal C}_{2}^{L}$ measured for different  times. 
As expected, with the decrease of $\Omega$, it is progressively increased until $\tau=149$ ns. 
Due to the experimental imperfections, the maximum of ${\cal C}_{2}^{L}$ is limited to 0.82. 

\begin{figure}[t]
  \centering
  \includegraphics[width=8.3cm]{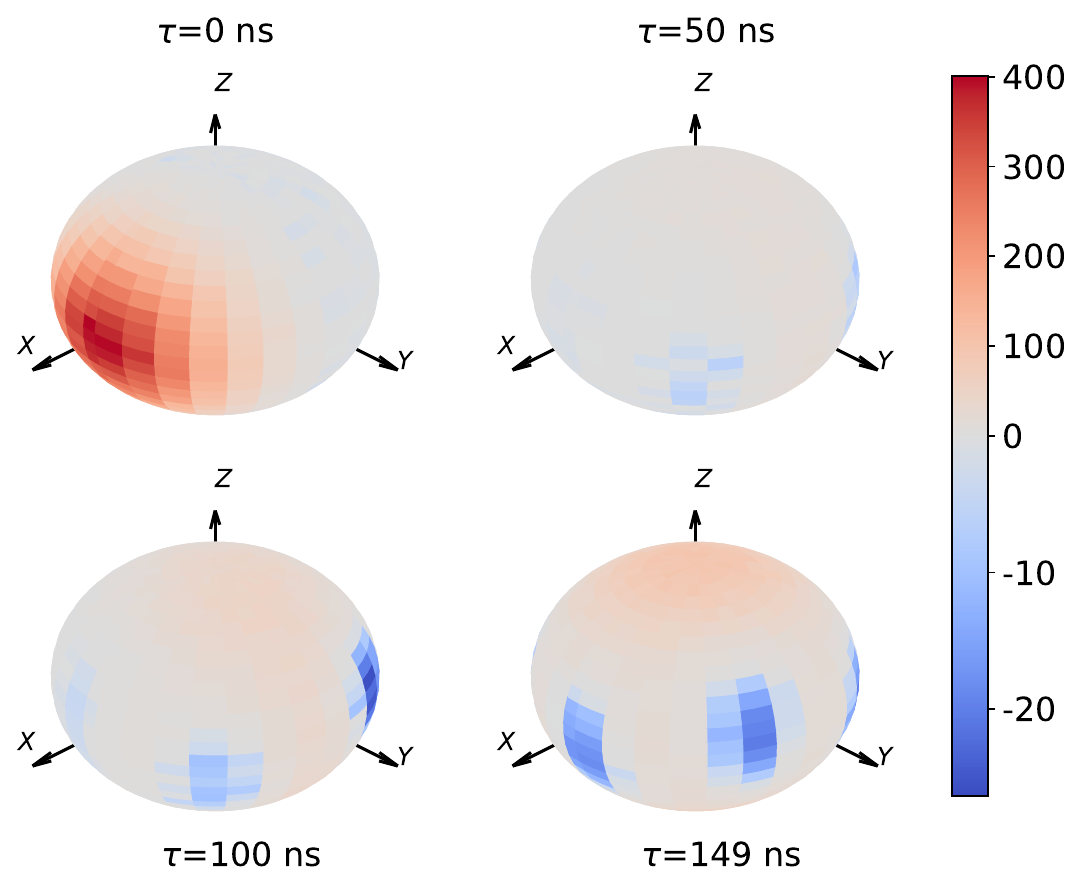} 
  \caption{Multi-qubit Wigner distributions measured for different evolution times. The Wigner function at $(\theta,\varphi )$ is obtained by performing a $\theta $-rotation on each qubit around the axis with an angle $\varphi -\pi /2$ to the x-direction on the equatorial plane, and then reading out its state.} 
  \label{fig4}
\end{figure}

The quantum coherence can be further revealed by the multi-qubit Wigner function, defined as \cite{Rundle2017, Song2019} 
\begin{eqnarray}
{\cal W}(\theta ,\varphi )={\rm Tr}\left[\rho U(\theta ,\varphi )\Pi U^{\dagger }(\theta ,\varphi )\right]\text{,}
\end{eqnarray}
where $U(\theta ,\varphi )=\exp \left[-i\theta/2  \left(\sin \varphi \sigma _{j}^{x}-\cos \varphi \sigma _{j}^{y} \right) \right]$ and $\Pi =\otimes_{j=1}^{6}(1-\sqrt{3}\sigma _{j}^{z})$. 
To measure the Wigner function at $(\theta,\varphi )$, the joint state readout is performed after a $\theta-$rotation around the axis with an angle $\varphi -\pi /2$ to the x-direction on the equatorial plane. 
The Wigner function reconstructed at $\tau =0$, 50, 100, and 149 ns are displayed in Fig. \ref{fig4}. 
The Wigner function corresponds to a quasi-probability distribution of the multi-qubit system on a unit sphere.
The regions where the values are negative reflect the quantum interference effects, and the negativity serves as a signature of multi-qubit cat states \cite{Song2019}. 
{
These results clearly verify that the two symmetry-breaking state components are superimposed with each other during the QPT. We note that this quantum-mechanical feature vanishes when the qubit number tends to infinity due to the linear scaling of the decoherence rates (see Sec. S7 of the Supplemental Material \cite{supp} for details), thereby distinguishing phase transitions of finite-size quantum many-body systems from those of infinite-size ones.}

{In conclusion, we have presented a demonstration of the SSB in the QPT of the multi-qubit LMG model, engineered with a superconducting processor.} 
The qubits are effectively coupled to each other through virtual photon exchange, mediated by a bus resonator dispersively coupled to the qubits. 
By tuning the competition between the external continuous driving and intra-qubit interaction, we can move the system from the $Z_{2}$-symmetric phase to the symmetry-breaking phase, where the two degenerate highest eigenstates of the governing Hamiltonian, superimposed with each other, form a GHZ-like cat state. 
We characterize the SSB extent with the average two-qubit longitudinal correlation. 
The quantum nature of the SSB phase is confirmed by the measured multi-qubit quantum correlation, as well as by the reconstructed multi-qubit Wigner function, which exhibits interference fringes with alternating positive and negative values. 
Besides fundamental interest, the demonstrated SSB process can be used for noise-insensitive many-body entanglement manipulation.
    
\begin{acknowledgments}
We thank Yi-Zhuang You at UC San Diego for the helpful discussion. 
This work was supported by the National Natural Science Foundation of China (Grants No. 12274080, No. 12474356, No. 12475015, No. 12204105, No. T2322030, No. T2121001, and No. 92065114), the Beiiing Nova Program (Grant No. 20220484121), and the Innovation Program for Quantum Science and Technology (Grants No. 2021ZD0300200 and No. 2021ZD0301800).
\end{acknowledgments}

\nocite{Facchi2002, Wang2008, Xu2023}

\bibliography{references}

\end{document}


\setcounter{secnumdepth}{3}
 
\title{Supplemental Material for \\ ``Experimental Demonstration of Spontaneous Symmetry Breaking with Emergent MultiQubit Entanglement"}

\author{Ri-Hua Zheng}
\thanks{These authors contribute equally to this work.}
\author{ Wen Ning}
\thanks{These authors contribute equally to this work.}\author{Jia-Hao L\"{u}} 
\author{Xue-Jia Yu}
\thanks{These authors contribute equally to this work.}

\author{Fan Wu}
\affiliation{Fujian Key Laboratory of Quantum Information and Quantum Optics, College of Physics and Information Engineering, Fuzhou University, Fuzhou, Fujian, 350108, China}

\author{ Cheng-Lin Deng }
\affiliation{Institute of Physics and Beijing National Laboratory for Condensed Matter Physics, Chinese Academy of Sciences, Beijing 100190, China}
\affiliation{CAS Center for Excellence in Topological Quantum Computation, University of Chinese Academy of Sciences, Beijing 100190, China}

\author{ Zhen-Biao Yang}\thanks{E-mail: zbyang@fzu.edu.cn}
\affiliation{Fujian Key Laboratory of Quantum Information and Quantum Optics, College of Physics and Information Engineering, Fuzhou University, Fuzhou, Fujian, 350108, China}
\affiliation{Hefei National Laboratory, Hefei 230088, China}

\author{Kai Xu}\thanks{E-mail: kaixu@iphy.ac.cn}
\author{ Dongning Zheng}
\author{ Heng Fan}
\affiliation{Institute of Physics and Beijing National Laboratory for Condensed Matter Physics, Chinese Academy of Sciences, Beijing 100190, China}
\affiliation{CAS Center for Excellence in Topological Quantum Computation, University of Chinese Academy of Sciences, Beijing 100190, China}
\affiliation{Hefei National Laboratory, Hefei 230088, China}

\author{Shi-Biao Zheng}\thanks{E-mail: t96034@fzu.edu.cn}
\affiliation{Fujian Key Laboratory of Quantum Information and Quantum Optics, College of Physics and Information Engineering, Fuzhou University, Fuzhou, Fujian, 350108, China}
\affiliation{Hefei National Laboratory, Hefei 230088, China}
\thanks{E-mail: zbyang@fzu.edu.cn}

\maketitle

\tableofcontents 

\newpage  

\section{Engineering the LMG model with competition of longitudinal couplings and transverse drivings} \label{H}
The $N$-qubit system Hamiltonian is (setting $\hbar=1$ hereafter)
\begin{eqnarray} \label{original_H}
H_1= \omega_{b} a^{\dagger} a +\sum_{j=1}^{N}\left[\omega \left|e \rangle_j\langle e\right|+\left(\xi_j \left|e \rangle_j\langle g\right| a + \frac{\Omega}{2}  e^{-i\omega_o t} \left|e \rangle_j\langle g\right|  +{\rm H.c.} \right) \right],
\end{eqnarray}
where $\omega_b/(2\pi)$, $\omega/(2\pi)$, and $\omega_o/(2\pi)$ are the frequencies of the resonator, the qubits, and the operating point, respectively. 
Additionally, $a$ is the annihilation operator of the resonator and $|g\rangle_j (|e\rangle_j)$ is the ground (first excited) state of $j$th qubit.
The qubit-resonator coupling is $\xi_j$ and the continuous microwaves own the same frequency $\omega_o/(2\pi)$ and amplitude $\Omega/2$.
When the qubit-resonator detuning is much larger than the coupling, i.e., $|\Delta|=|\omega_o-\omega_b| \gg \xi_j$, and the resonator field is initially in the vacuum state, the effective Hamiltonian in the interaction frame becomes \cite{Zheng2000},
\begin{eqnarray}
H_1'=  \sum_{i,j}^{N} \frac{\xi_i\xi_j}{\Delta} \sigma^+_j \sigma^-_j   + \frac{\Omega}{2}  \sigma^x_j + \frac{ \omega-\omega_o}{2} \sigma^z_j,
\end{eqnarray}
with $\sigma_j^{\pm} = |e\rangle_j \langle g| \ (|g\rangle_j \langle e|)$ and $\sigma^x_j=\sigma^+_j+\sigma^-_j$.
For simplicity, we set $\xi_j=\xi$ and $\lambda=-\xi^2/\Delta$ ($\Delta<0$ in this experiment), yielding
\begin{eqnarray}
H_1'=  -\lambda S_+ S_-  + \Omega S_x +(\omega-\omega_o)S_z,
\end{eqnarray}
where $S_\pm=\sum_j^N \sigma^\pm_j$ and $S_{x,y,z}=\sum_j^N \sigma^{x,y,z}_j/2$.
If the state of the $N$-qubit system is symmetrically expanded as $|J,m\rangle$ ($J$ denotes the total angular momentum quantum number that is equal to $N/2$ for the symmetric case and $m=-J,-J+1...,J$), we have
\begin{align}
S_+|J,m \rangle & = \sqrt{(J-m)(J+m+1)}|J,m+1\rangle \label{eq:Splus}, \\
S_-|J,m\rangle & = \sqrt{(J-m+1)(J+m)}|J,m-1\rangle \label{eq:Sminus}, \\
S_z |J,m\rangle& =\frac{1}{2}\left(S_{+} S_{-}-S_{-} S_{+}\right)|J,m\rangle=m |J,m\rangle\label{eq:Sz},
\end{align}
such that
\begin{eqnarray}
S_+S_-|J,m \rangle=(-m^2+m+J^2+J)|J,m\rangle=(-S_z^2+S_z+J^2+J)|J,m\rangle,
\end{eqnarray}
inducing
\begin{eqnarray}
H_1'=  -\lambda (-S_z^2+S_z)  + \Omega S_x +(\omega-\omega_o)S_z,
\end{eqnarray}
with ignoring the trivial constant term $J^2+J$.
By adjusting $\omega_o=\omega-\lambda$, the effective Hamiltonian is given by 
\begin{eqnarray} \label{eff_H}
H_{\rm eff}=  \Omega S_x +\lambda S_z^2,
\end{eqnarray}
as $-H$ in the main text, where 
\begin{eqnarray} 
H=  -\Omega S_x -\lambda S_z^2. \label{eq:LMG}
\end{eqnarray}

To clarify the underlying physics of $H$ more clearly, we perform the unitary transformation $e^{-i\Omega tS_{x}}$, under which the
qubit-qubit swapping interactions are described by the Hamiltonian
\begin{eqnarray}
H_{\rm swap} =\frac{\lambda }{2}\big[S_{x}^{2}-\cos (2\Omega t)(S_{z}^{2}-S_{y}^{2}) +\sin (2\Omega t)(S_{z}S_{y}+S_{y}S_{z})\big].  
\end{eqnarray}
Here we have discarded the trivial constant term $\lambda J(J+1)/2$. 
Under the condition $\lambda \ll \Omega $, the fast-oscillating terms can be discarded, and $H_{\rm swap}$ reduces to $H_{\rm swap}\simeq  \lambda S_{x}^{2}/2$, which commutes with the transverse driving Hamiltonian and does not affect the state $\left\vert X\right\rangle$. 
This phenomenon can also be understood in terms of dynamical decoupling \cite{Guo2018} or Zeno dynamics \cite{Facchi2002, Wang2008} produced by the driving, which blocks the transition between $\left\vert \pm _{j}\right\rangle $, thereby freezing the entanglement dynamics.

The behavior in the regime $\Omega \ll \lambda $ can be understood by performing the transformation $e^{-i\lambda tS_{z}^{2}}$, under which the driving Hamiltonian becomes 
\begin{eqnarray}
H_{\rm driv}=-\stackrel{J-1}{ \mathrel{\mathop{\sum }\limits_{m=-J}}}\Omega _{m}e^{i\lambda _{m}t}\left\vert J,m\right\rangle \left\langle J,m+1\right\vert +{\rm H.c.}, 
\end{eqnarray}
where $\lambda _{m}=(2m+1)\lambda $, $\Omega _{m}=\Omega \sqrt{(J-m)(J+m+1)}$, and $\left\vert J,m\right\rangle $ is the eigenstate of $S_{z}$ with eigenvalue $m$. When $\Omega \ll \lambda $, the system can't make transitions between the eigenstates with different $m$, $H_{\rm driv}$ can be replaced by the effective Hamiltonian
\begin{eqnarray}
H_{\rm driv}^{\rm eff}=\stackrel{J}{
\mathrel{\mathop{\sum }\limits_{m=-J}}}\eta _{m}\left\vert J,m\right\rangle \left\langle J,m\right\vert , 
\end{eqnarray}
where $\eta _{J}=\eta _{-J}=-2J\Omega ^{2}/[(2J-1)\lambda ]$, and $\eta_{m}=\Omega _{m}^{2}/\lambda _{m}-\Omega _{m-1}^{2}/\lambda _{m-1}$ for $m\neq J,-J$. This implies that the transverse driving only slightly shifts the energy levels of the eigenstates.
The most remarkable feature of the system in the strong-coupling regime is that it has two degenerate ground states, $\left\vert J,J\right\rangle=\otimes _{j=1}^{N}\left\vert e_{j}\right\rangle $ and $\left\vert J,-J\right\rangle =\otimes _{j=1}^{N}\left\vert g_{j}\right\rangle $, each breaking the system's $Z_{2}$ symmetry.

\section{Quantum phase transitions in ideal LMG models}

\begin{figure}[htbp]
  \centering
  \includegraphics[width=16cm]{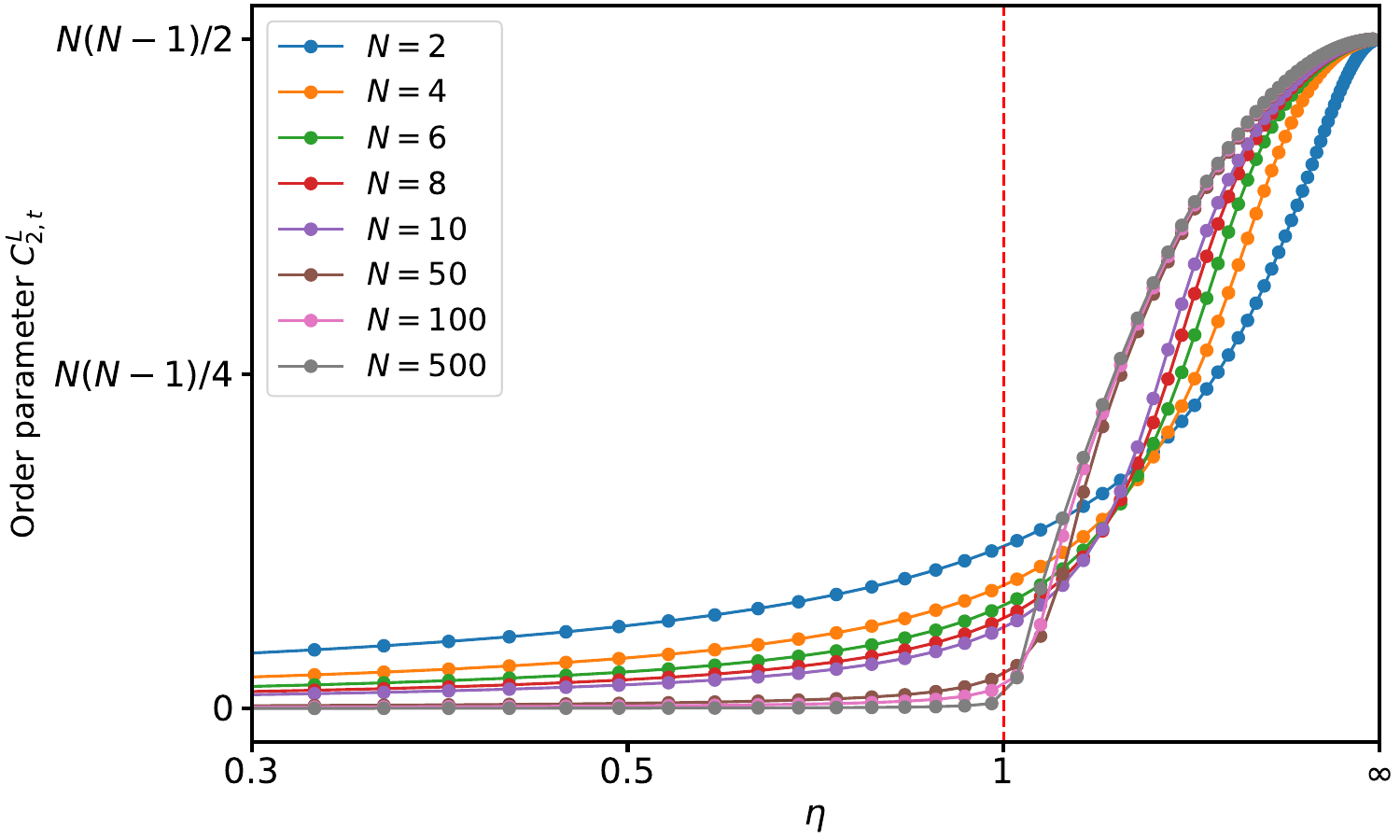} 
  \caption{\textbf{Critical behaviors for the ideal LMG model with different system sizes.} For $N \le 10$, the results are derived from direct simulation of the ground state dynamics in the Hilbert space of $N$ qubits. 
  For $N > 10$, the simulation are conducted within the symmetric-Dicke-state subspace.
  It is quite evident that as $N$ increases, the first derivative of the order parameter $C_{2,t}^{L}$  becomes increasingly discontinuous at the critical point $\eta=1$ (marked by the red dashed line), and the phase transition phenomenon in the system becomes more and more pronounced.
  }
  \label{largeN}
\end{figure}

Here the critical behaviors for the ideal LMG model with different system sizes are investigated by the numerical simulation.
We numerically solve for the ground state of the Hamiltonian in Eq. (\ref{eq:LMG}) with respect to different $\eta=N\lambda/\Omega$ values, and use the ground state to compute the order parameter, thereby obtaining the static phase diagram of the LMG model shown in Fig. \ref{largeN}.
Note that the results for $N \le 10$ are obtained through direct simulation of the ground state dynamics in the $N$-qubit Hilbert space. 
While the results for $N>10$ are obtained by projecting the system space onto the symmetric-Dicke-state subspace by Eqs. (\ref{eq:Splus},\ref{eq:Sminus},\ref{eq:Sz}) before simulation to reduce the computational cost.

We note that the phase transition of this model can be measured by the order parameter, defined as the total two-qubit longitudinal correlation,
\begin{eqnarray}
C_{2,t}^{L}=\sum_{k>j}^{N}\sum_{j=1}^{N-1}\langle \sigma_{j}^{z}\sigma_{k}^{z} \rangle.
\end{eqnarray}
This correlation is equal to 0 when $\eta \to 0$, and tends to $N(N-1)/2$ for $\eta \to \infty$. Figure. \ref{largeN} displays $C_{2,t}^{L}$ as a function of $\eta$, which reveals that the changing rate of $C_{2,t}^{L}$ at the point $\eta=1$ depends upon the number of the qubits. For the 6-qubit system, $C_{2,t}^{L}$ does not show a sudden change at this point, but the GHZ entanglement still emerges when $\eta \gg 1$ as a consequence of SSB.

\section{Experiment parameters}

\subsection{Parameters of the qubits}

\begin{table}
\begin{center}
\begin{tabular}{ccccccc}
  \hline
  \hline
  & \hspace{1em} $Q_1$ \hspace{1em}  & \hspace{1em} $Q_2$ \hspace{1em}  & \hspace{1em} $Q_3$ \hspace{1em}  & \hspace{1em} $Q_4$ \hspace{1em}  & \hspace{1em} $Q_5$ \hspace{1em}  & \hspace{1em} $Q_6$ \hspace{1em}  \\
  \hline
  \hspace{2em} $\omega_i$ (2$\pi$ GHz) \hspace{2em} &  5.25    &   5.146   &   5.19   &   5.657    &   5.3   &  5.08    \\
  $F_0$ &   0.95  &   0.94   &  0.99    &   0.99   &  0.98    &   0.96   \\
  $F_1$ &   0.93  &   0.92   &  0.88    &   0.87   &  0.94    &   0.90   \\
  $\omega_r$ (2$\pi$ GHz) &   6.6944   &   6.8109   &  6.6376     &   6.6195   &  6.6966    &   6.5111   \\
  $\omega_s$ (2$\pi$ GHz) &   5.780   &   5.822   &  5.715    &   5.730   &   5.77  &   5.748   \\
  $T_1$ $(\mu s)$ &   9.5  &   14.2 &   16.4  &   9.0   &   13.3  &    31.6  \\
  $T_2$ $(\mu s)$ &   4.1  &    4.3  &   3.2   &  4.6    &    2.2  &    3.8  \\
  $T_2^{\rm SE}$ $(\mu s)$ &  10.1    &   13.6   &   8.4   &   9.8    &   6.8   &   18.4   \\
  $T_2^{\rm c}$ $(\mu s)$ &  8.8    &   12.8   &   9.6  &   9.3    &   11.3   &   24.4   \\
  $\xi_j$ (2$\pi$ MHz) &   19.56   &   19.78   &   15.98   &   19.24   &   19.88   &   14.51   \\
  \hline
  \hline
\end{tabular}
\end{center}
  \caption{\textbf{Characteristics of the qubits.} 
  The idle frequencies of $Q_j$ ($j=1,2,..,6$) are marked by $\omega_{i}/(2\pi)$, where single-qubit rotation pulses and measurements are applied. 
  The probability of detecting the qubit in $\vert g\rangle$ ($\vert e\rangle$) when it is prepared in $\vert g\rangle$ ($\vert e\rangle$) state is indicated by $F_{g}$ ($F_{e}$).
  Each qubit's state is readout by the dispersion effect between the qubit and the corresponding readout resonator, whose frequency is $\omega_r/(2\pi)$.
  The highest frequency point at which each qubit can be tuned is labeled as $\omega_s/(2\pi)$.
  For the decoherence performance, $T_{1}$ represents the energy relaxation time, $T_2$ is the Ramsey dephasing time (Gaussian fit), $T_2^{\rm SE}$ denotes the spin-echo dephasing time (exponential fit), and $T_2^{\rm c}$ refers to the continuous-driving dephasing time (exponential fit), all measured at the working frequency of 5.6895 GHz. 
  The coupling strength between $Q_j$ and the bus resonator is denoted by $\xi_j$.}
  \label{paras}
\end{table}

The performance characterization of the qubits is listed in Table \ref{paras}. 
For technical details about the superconducting circuits, e.g., see Refs. \cite{Song2017, Song2019, Xu2020, Xu2023}, which report similar qubits' parameters and control methods to our experiment.

\subsection{Measurement of the frequency $\omega_b$ of the bus resonator}

To measure the frequency of the resonator, we drive the resonator with microwaves using the xy line of $Q_6$ (Because the xy line of $Q_6$ is physically close to the resonator, see Fig. 1a in the main text). 
This driving Hamiltonian is
\begin{eqnarray}
H_d=\Omega_b e^{i \omega_b't} a + {\rm H.c.},
\end{eqnarray}
where $\Omega_b$ and $\omega_b'$ are the driving amplitude and frequency, respectively.
When this driving frequency $\omega_b'$ is near the real resonator frequency $\omega_b$, the resonator will accumulate photons.

Then the photon number accumulation of the resonator can be readout by utilizing another qubit $Q_2$ on resonance with the resonator, whose dynamics is described by 
\begin{eqnarray}
H_r=\xi_2 |g\rangle \langle e| a^\dag + {\rm H.c.}.
\end{eqnarray}
The reason for choosing $Q_2$ is that it has the highest tunable frequency. 
Note that in this circuit (Fig. 1a in the main text), most of the qubits' frequencies are unable to reach the resonator frequency $\omega_b$.
This means that most qubits cannot resonate with the resonator.
After the qubit-resonator interaction of a specific time, a portion of the resonator's photons are converted into the excitation of $Q_2$. 
In this way, we can infer whether there are photons in the resonator by reading out $Q_2$.

\begin{figure}[htbp]
  \centering
  \includegraphics[width=16cm]{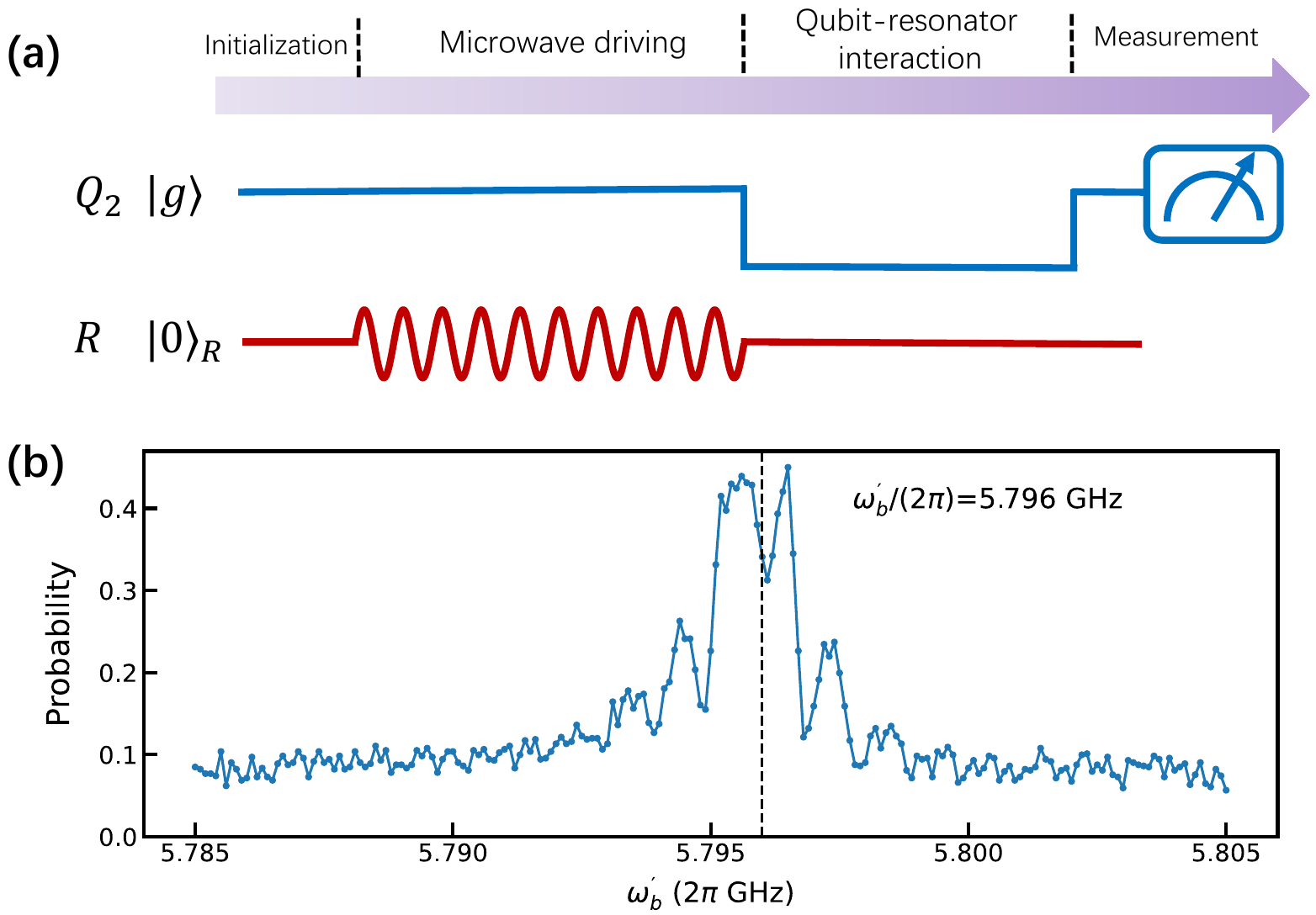} 
  \caption{\textbf{Measurement of the resonator frequency.} (a) Pulses sequence.
  Firstly, the qubit and resonator $R$ are initially prepared in the ground state $|g\rangle$ and the vacuum state $|0\rangle_R$, respectively. 
  Secondly, we drive the resonator using the xy line of $Q_6$. 
  Thirdly, the z line of $Q_2$ is tuned by a square wave to resonate with the resonator. 
  After the qubit-resonator interaction, we can measure $Q_2$ to determine if the resonator is excited by the microwave driving $\Omega_b e^{i \omega_b't}$.
  (b) Measured resonator signal $P_{|e\rangle}$ versus the microwave driving frequency $\omega_b'$. 
  When the microwave frequency $\omega_b'$ approaches the real frequency of the resonator, $\omega_b$, a distinct series of peaks can be seen. 
  The center of these peaks, marked with a dotted line, is the measured resonator frequency, $\omega_b/(2\pi) = 5.796$ GHz.
  }
  \label{resonator}
\end{figure}

The corresponding pulse sequence and measured data are shown in Fig. \ref{resonator}, from which we can see that the frequency of the resonator is $\omega_b/(2\pi) = 5.796$ GHz.

\subsection{Mapping the qubit's frequency to the z-line square pulse amplitude} \label{Sec_f2zpa}

\begin{figure}[htbp]
  \centering
  \includegraphics[width=16cm]{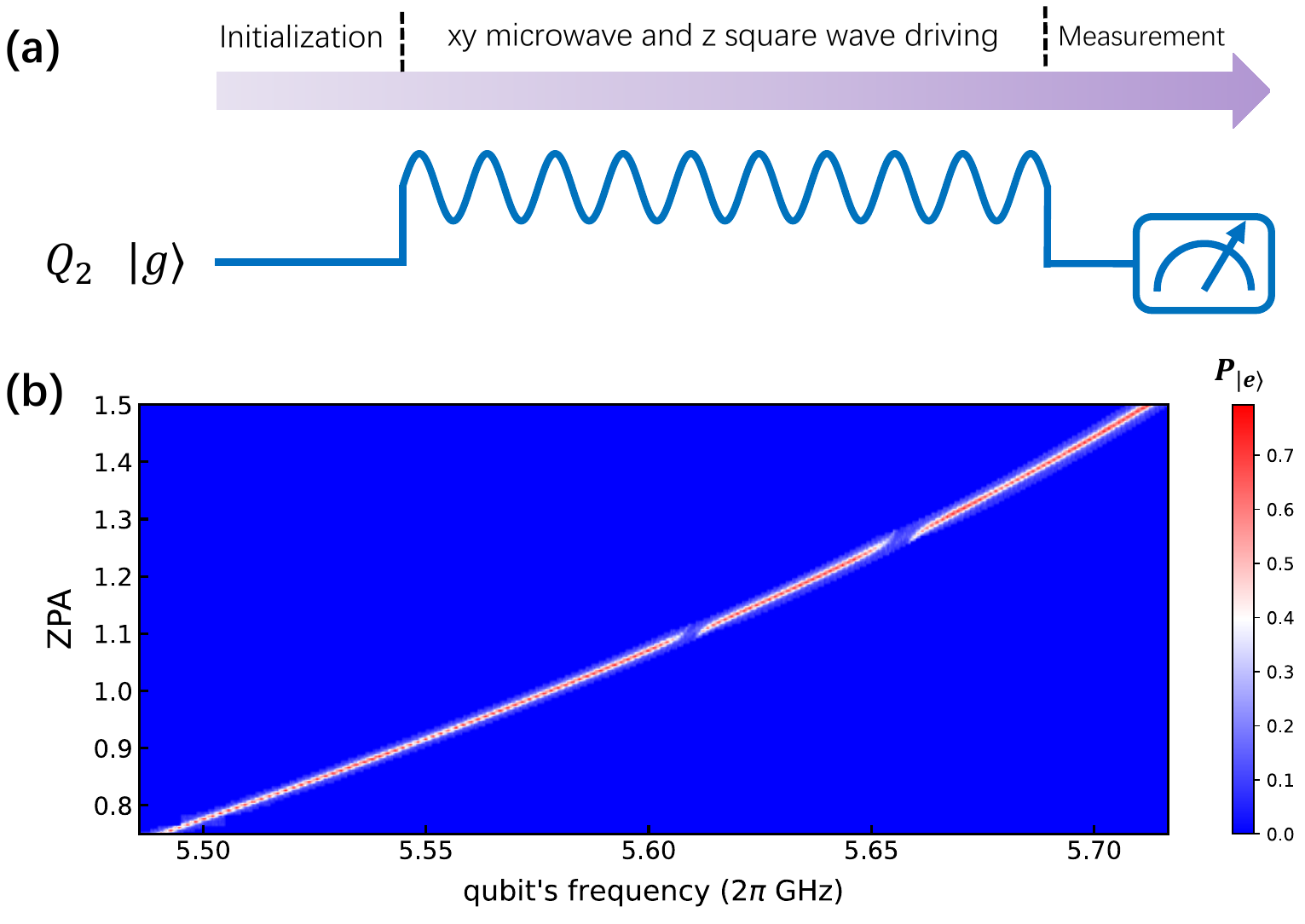} 
  \caption{\textbf{Mapping diagram between the frequency of $\bm {Q_2}$ and the ZPA.} (a) Pulses sequence.
  The qubit $Q_2$ is initially prepared in the ground states $|g\rangle$. 
  Then, we simultaneously drive $Q_2$ with the xy-line microwave and z-line square wave and finally measure the qubit.
  When the frequency of the xy-line microwave corresponds to the ZPA of the z pulse, $Q_2$ will be excited.
  (b) Measured population versus the qubit's frequency and the ZPA. 
  The excitation signal of $Q_2$ is demonstrated in the diagram through an arresting bright curve, depicting the mapping relationship between the frequency of $Q_2$ and ZPA.
  }
  \label{spec}
\end{figure}

In the experiment, we tune the qubit's frequency by driving a square wave through the z-line of the qubit.
Therefore, a mapping of the relationship between the qubit's frequency and the z pulse amplitude (ZPA) is necessary.
We can simultaneously send a microwave with a variable frequency to the qubit's xy line and a square wave with a specific ZPA to the qubit's z line.
If the frequency of the xy-line microwave is equal to the frequency of the qubit driven by the z pulse with a specific ZPA, then one can see excitation in the signal.
These pulse sequences are shown in Fig. \ref{spec}(a), and in Fig. \ref{spec}(b) we exhibit a detailed mapping diagram of ZPA versus the frequency of $Q_2$.
One can fit the relationship between the frequency of $ {Q_2}$ and the ZPA through Fig. \ref{spec}(b), and it's similar for the other qubits as well.

\subsection{Measurement of the qubit-resonator couplings}

\begin{figure}[htbp]
  \centering
  \includegraphics[width=16cm]{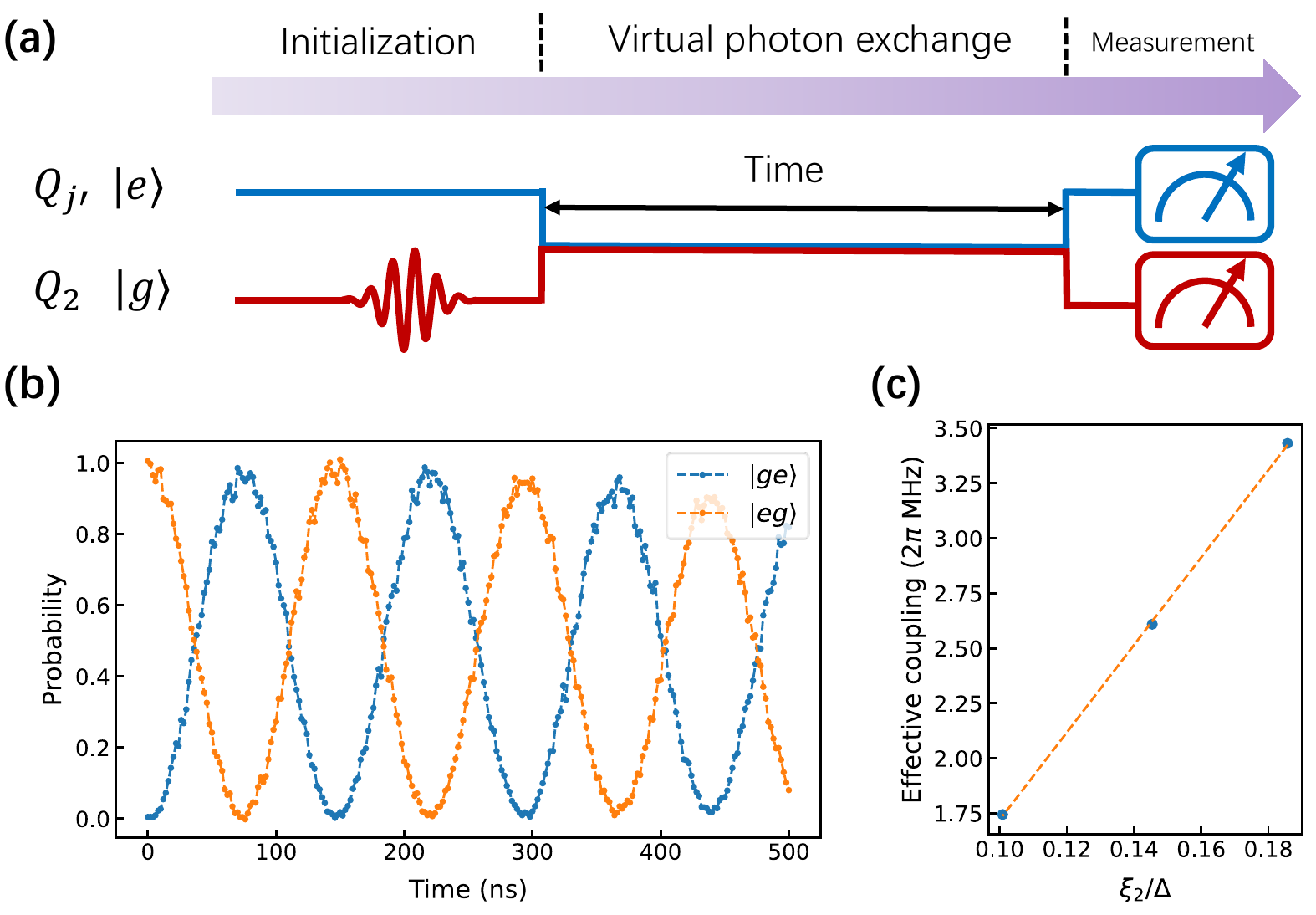} 
  \caption{\textbf{Measurement of the qubit-resonator coupling strength.} (a) Pulses sequence.
  The qubits $Q_2$ and $Q_{j}'$ are initially prepared in the ground state $|g\rangle$ and the first excited state $|e\rangle$, respectively. 
  Then, we simultaneously bias $Q_2$ and $Q_{j'}$ to the same operating point $\omega_{\rm op}$ by the z-line square waves with specific ZPAs. 
  The detuning between this operating point and the resonator frequency is $\Delta'=\omega_{\rm op}-\omega_{b}$.
  This virtual photon exchange process will lead to the population oscillation between $|eg\rangle$ and $|ge\rangle$ with the effective coupling $\xi_{j'}\xi_2/\Delta'+b$, where $b$ is the fixed XX-type coupling strength.
  (b) The virtual photon exchange process of $Q_2$ and $Q_5$ at operating point $\omega_{\rm op}/(2\pi)=5.6895$ GHz, as an example.
  Half of this population oscillation frequency is the effective coupling strength $\xi_{5}\xi_2/\Delta'+b$.
  (c) Effective coupling strength versus different $\xi_2/\Delta'$, measured at operating point $\omega_{\rm op}/(2\pi)=5.6895$, 5.6600, 5.6000 GHz. 
  The slope of this line represents the qubit-resonator coupling strength $\xi_5=2\pi \times 19.88$ MHz and the intercept represents the fixed XX-type coupling strength $b=2\pi \times-0.27$ MHz.
  }
  \label{cal_coupling}
\end{figure}

In the superconducting circuit (Fig. 1a in the main text), the qubit $Q_j$ and resonator are capacitatively coupled with strength $\xi_j$.  
The coupling of each qubit to the resonator is indeed not uniform. 
We show here how to measure the coupling strength $\xi_j$ between the qubits and the resonator. 
For qubit $Q_2$, we can obtain the coupling strength $\xi_2=2\pi \times 19.78$ MHz by resonating it with the resonator and by measuring the vibration frequency of the qubit population. 
For other qubits, however, their highest tunable frequencies can not reach the resonator frequency $\omega_b/(2\pi)=5.796$ GHz.
So we bias one of the other qubits and $Q_2$ simultaneously to an operating point $\omega_{\rm op}$ more than $2\pi \times$ 100 MHz away from the resonator frequency $\omega_b$. 
These two qubits will undergo a second-order virtual photon exchange process (see Sec. \ref{H}) with swapping rate $\xi_{j'}\xi_2/(\omega_{\rm op}-\omega_{b})$ ($j' \neq 2$) if the initial state is $|eg\rangle$ or $|ge\rangle$.
It is worth mentioning that there is XX-type coupling between each pair of qubits (forming as $b |eg\rangle \langle ge| a^\dag + {\rm H.c.}$) and $b$ is different for each pair of qubits. 
So we should measure the virtual photon exchange at several operating points (at least two) to get $\xi_{j'}$ and $b$, see Fig. \ref{cal_coupling}(c).
In Figs. \ref{cal_coupling}(a) and \ref{cal_coupling}(b), we show the corresponding pulse sequence and the population oscillation of $Q_2$ and $Q_5$, respectively.

\section{Realization of the quasi-adiabatic process}

As mentioned in the main text, the experiment starts by tuning all these qubits to their idle frequencies and preparing each qubit in state $\left\vert +_{j}\right\rangle =\frac{1}{\sqrt{2}}(\left\vert g_{j}\right\rangle +\left\vert e_{j}\right\rangle )$ by applying a $\pi /2$ pulse.
This is followed by the application of a continuous pulse, incorporating the transverse drivings with the longitudinal couplings.
The phase transition is realized by a quasi-adiabatic process, during which the driving strengths are configured as $\Omega (\tau )=\Omega^0 \exp(-\tau/t_f)$, where $\Omega^0/(2\pi)=40$ MHz, $t_f=60$ ns, and $\tau$ represents the time, as shown in Fig. 2a in the main text. 
After a preset time, the transverse drives are switched off, following which each of the qubits is biased back to its idle frequency for state readout.

To realize quasi-adiabatic dynamics and readout the final state, we need to calibrate the crosstalk of the qubit's xy and z lines and cancel out the random phases of the initial states, the continuous microwaves, and the rotation pulses.


\subsection{Calibration of the Z crosstalk}\label{Sec_Zcross}
In the experiment, the qubits' z lines will interfere with each other due to the physical distance.
This means that when a qubit receives a z-line square wave with a certain ZPA, there is also a small z-line square wave input (by the Z crosstalk) for the other qubits, which causes the mapping of the previous qubit frequency versus the ZPA to fail in the Sec. \ref{Sec_f2zpa}, i.e., the qubits cannot be tuned to the desired frequencies.

\begin{figure}[htbp]
  \centering
  \includegraphics[width=16cm]{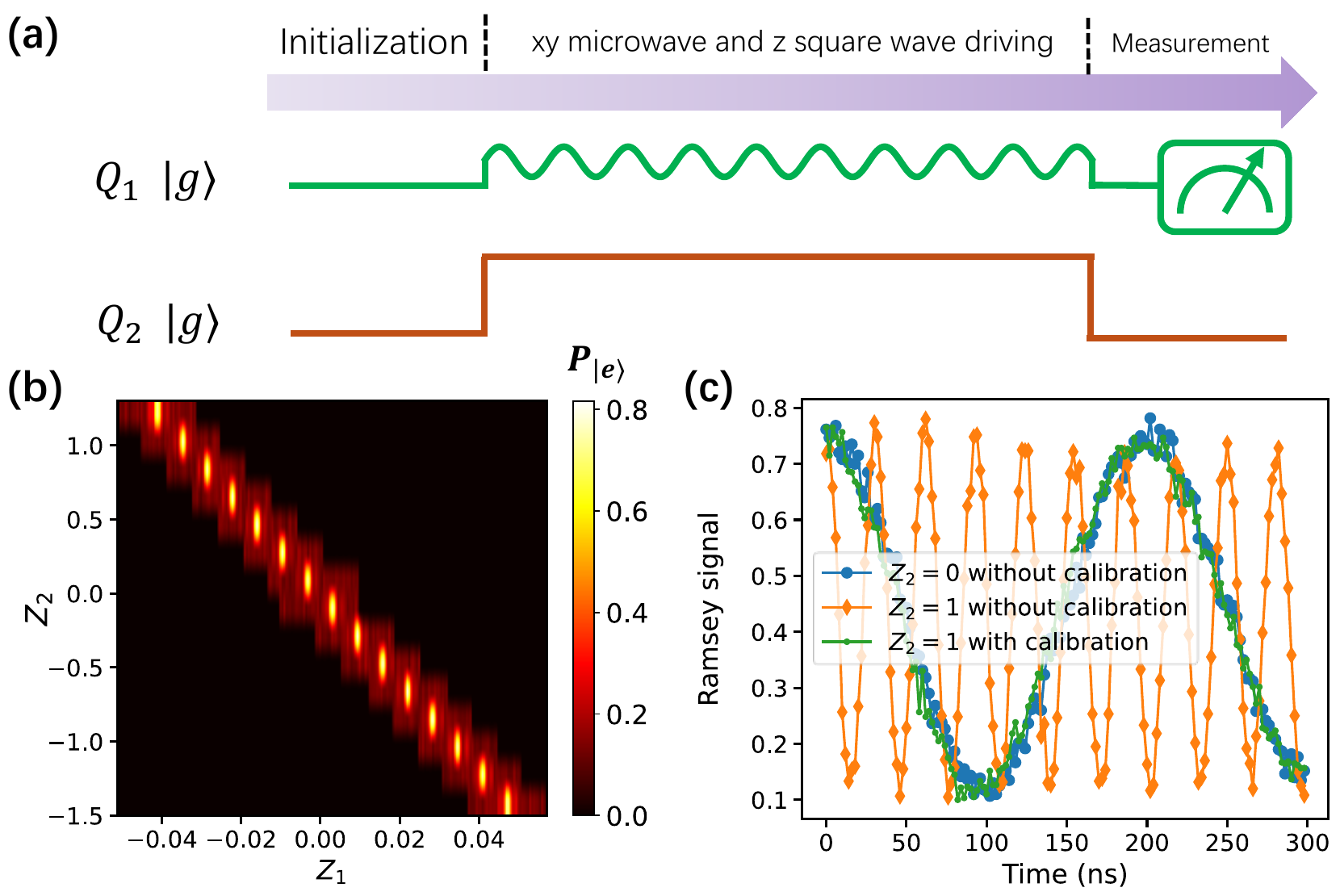} 
  \caption{\textbf{Calibration of the Z crosstalk.} (a) Pulse sequence.
  For example, here we calibrate the Z crosstalk of $Q_2$ against $Q_1$.
  The qubit $Q_1$ is initially prepared in the ground state $|g\rangle$.
  A specific z-line square wave with ZPA $Z_2$ is applied to $Q_2$, while a z-line square wave with smaller ZPA $Z_1$ is applied to $Q_1$ to compensate for the effect from $Q_2$'s z line on $Q_1$. 
  In the meanwhile, we add a xy-line microwave whose frequency is equal to the idle frequency of $Q_1$ to excite $Q_1$. 
  If the compensation of $Z_1$ is successful, $Q_1$ can be excited. 
  (b) Measured excited population versus $Z_1$ and $Z_2$. 
  The bright line marks the linear relationship between $Z_1$ and $Z_2$, from which it can be seen that $Z_2=1$ of $Q_2's$ z-line square wave will approximately induce a square wave with a ZPA of -0.04 on $Q_1$.
  (c) Ramsey signal versus time. Here three scenarios are considered: $Z_2=0$ without calibration, $Z_2=1$ without calibration, and $Z_2=1$ with calibration. 
  From this, one can see the influence of the Z crosstalk on $Q_1$'s frequency and the effectiveness of calibration of the Z crosstalk.
  }
  \label{Zcross}
\end{figure}

Calibration of the Z crosstalk, taking $Q_1$ and $Q_2$ as an example, can be achieved by inputting a z-line square wave with a specific ZPA $Z_2$ to $Q_2$, and then inputting a small z-line square wave with ZPA $Z_1$ to $Q_1$ to compensate for the influence of $Q_2$'s z line on it. 
At the same time, we input a microwave to the xy-line of $Q_1$ with a frequency equal to its idle frequency [See Fig. \ref{Zcross}(a)]. 
If the crosstalk of the z line is compensated, the frequency of $Q_1$ will not change, so $Q_1$ will be excited by the xy-line microwave.
Therefore, scanning a two-dimensional plot of the excitation $P_{|e\rangle}$ versus $Z_1$ and $Z_2$ can obtain the relationship between $Z_1$ and $Z_2$ in Fig. \ref{Zcross}(b). 
In this way, for different $Z_2$ of $Q_2$'s z-line square wave, one can know the needed $Z_1$ of $Q_1$'s z-line square wave to calibrate the corresponding Z crosstalk.

To check the calibration of the Z crosstalk of $Q_2 \to Q_1$, a Ramsey interferometry can be performed on $Q_1$ to infer the shift of $Q_1$'s frequency.
Specifically, we add two $X/2$ pulses to $Q_1$, with a resting time $t$ in between. 
The phase difference between these two $X/2$ pulses is $f_rt$, where the fringe frequency $f_r/(2\pi)$=5 MHz.
When the frequency of $Q_1$ is constant, one will obtain a Ramsey signal with frequency $f_r/(2\pi)$. 
But when the frequency of $Q_1$ changes (due to the Z crosstalk), the resulting Ramsey signal frequency changes with it. 
In this way, we can judge whether $Q_1$ is affected by the Z crosstalk.
In Fig. \ref{Zcross}(b) we consider three scenarios: (i) $Z_2=0$ without calibration, (ii) $Z_2=1$ without calibration, and (iii) $Z_2=1$ with calibration.
The difference between (i) and (ii) shows that the Z crosstalk has a considerable impact on $Q_1$'s frequency and the validity of this calibration can be seen from the consistency of (i) and (iii).

The Z crosstalk of $Q_1 \to Q_2$ can be calibrated similarly and further generalized to any two pairs of qubits.

\subsection{Calibration of the XY crosstalk}\label{Sec_cross}

\begin{figure}[htbp]
  \centering
  \includegraphics[width=16cm]{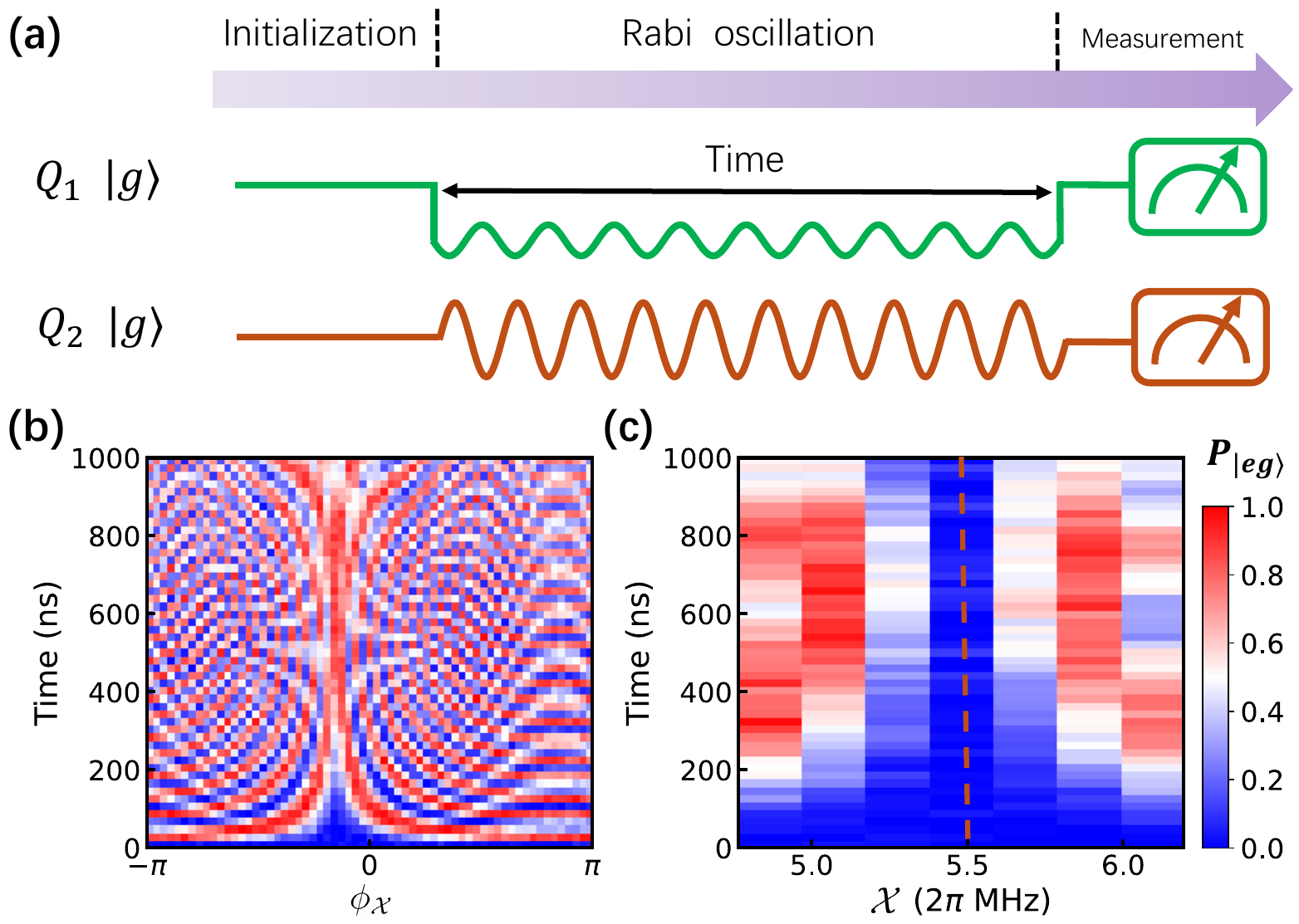} 
  \caption{\textbf{Calibration of the XY crosstalk.} (a) Pulse sequence.
  As an example, the qubits $Q_1$ and $Q_2$ are initially prepared in the ground state $|gg\rangle$. 
  Then, $Q_1$ is solely biased to the operating point $\omega_o/(2\pi)=5.6895$ MHz by the z-line square wave with a specific ZPA and we simultaneously input two xy-line microwaves, $B=2\pi \times$20 MHz and ${\cal X} e^{i\phi_{\cal X}}$ [both with frequency $\omega_o/(2\pi)$] to $Q_1$ and $Q_2$, respectively.
  The single excitation of $Q_1$, $P_{|eg\rangle}$, will be measured to construct the Rabi oscillation signal to deduce if ${\cal X} e^{i\phi_{\cal X}}$ can cancel out the XY crosstalk on $Q_1$ caused by $Q_2$'s xy-line microwaves.
  (b) Single-excitation population $P_{|eg\rangle}$ versus $\phi_{\cal X}$ and Rabi oscillation time. 
  It can be seen that the Rabi oscillation signal has the smallest frequency at $\varphi_{\cal X}=-0.5$.
  So we can fix the microwave phase at $\phi_{\cal X}=-0.5$ and subsequently scan the microwave amplitude $\cal X$.
  (c) Single-excitation population $P_{|eg\rangle}$ versus ${\cal X}$ and Rabi oscillation time. 
  After fixing the microwave phase $\phi_{\cal X}=-0.5$, the frequency of the Rabi vibration signal is minimum at the amplitude ${\cal X}=2\pi \times 5.5$ MHz.
  }
  \label{XYcross}
\end{figure}

Similar to the Z crosstalk between qubits, there is also crosstalk between xy-line drivings of qubits.
Taking $Q_1$ and $Q_2$ as an example, the xy-line microwave $A e^{i \phi_A}$ input to $Q_1$ actually has an effect on $Q_2$ as well, quantified as $a_{2 1} A e^{i\left(\phi_{21}+\phi_A\right)}$, where $a_{2 1}$ and $\phi_{21}$ are the amplitude and phase coefficients, respectively.
Homoplastically, the xy-line microwave of $Q_2$, $B e^{i\phi_B}$, have an effect on $Q_1$ as $a_{1 2}B e^{i\left(\phi_{12}+\phi_B\right)}$, so the actual microwaves received by $Q_1$ and $Q_2$ are
\begin{eqnarray}
\left(\begin{array}{c}
A e^{i \phi_A}+a_{12} B e^{i\left(\phi_{12}+\phi_B\right)} \\
a_{21} A e^{i\left(\phi_{21}+\phi_A\right)}+B e^{i \phi_B}
\end{array}\right)=M_{1,2}\left(\begin{array}{l}
A e^{i \phi_A} \\
B e^{i \phi_B}
\end{array}\right),
\end{eqnarray}
where
\begin{eqnarray}
M_{1,2}=\left(\begin{array}{cc}
1 & a_{12} e^{i \phi_{12}} \\
a_{21} e^{i \phi_{21}} & 1
\end{array}\right).
\end{eqnarray}
If one wants the xy-line signals received by $Q_1$ and $Q_2$ to be ideal as $(Ae^{i\phi_A}, Be^{i\phi_B})^T$, the input xy-line signals for $Q_1$ and $Q_2$ should be
\begin{eqnarray}
\left(M_{1,2} \right) ^{-1}\left(\begin{array}{l}
A e^{i \phi_A} \\
B e^{i \phi_B}
\end{array}\right).
\end{eqnarray}
For the measurement of the upper triangle of $M_{1, 2}$, $a_{12} e^{i \phi_{12}}$,
we bias $Q_1$ to the operating point $\omega_o$, and input a xy-line microwave with frequency $\omega_o/(2\pi)$ and amplitude $B$ to $Q_2$ (at this time, $Q_2$ is still at the idle frequency and will not be excited). 
At the same time, we input a microwave ${\cal X} e^{i\phi_{\cal X}}$ with frequency $\omega_o/(2\pi)$ to $Q_1$ to offset the effect of $Q_2$'s xy-line microwave on it [see Fig. \ref{XYcross}(a)].
The received xy-line microwave of $Q_1$ is now ${\cal X} e^{i\phi_{\cal X}}+a_{12} B e^{i \phi_{12}}$, which induces a zero-frequency Rabi signal at point $(\phi_{\cal X},{\cal X})=(\phi_{12}+\pi,a_{12} B) $.
Since each Rabi signal has a timeline, one can only scan one more parameter ($\phi_{\cal X}$ or $\cal{X}$) to find this minimum frequency point through two local optimizations.
Mathematically, it can be proven that the minimum value of the Rabi signal frequency on the phase axis $\phi_{\cal X}$ should be scanned first, before scanning the amplitude axis $\cal{X}$. 
This approach ensures that the point crosswise found by two local optimizations are global optimization points, while the reverse is not true.
The measured Rabi oscillation signals are shown in Figs. \ref{XYcross}(b) and \ref{XYcross}(c), where one can deduce
$(\phi_{\cal X},{\cal X})=(-0.5, 2\pi \times 5.5 \ {\rm MHz})$ and therefore $(\phi_{12},a_{12}) =(-3.64, 0.275)$.
In this way, the upper triangular part of the $M_{1,2}$ matrix can be measured, while the lower triangular part can be measured symmetrically, which can be further extended to the $M_{j,k}$ matrix of any two qubits $Q_j$ and $Q_k$. 
Combining these $M_{i,j}$ matrices can be represented as 
\begin{eqnarray}
{\cal M}=\left(\begin{array}{cccc}
1 & a_{12} e^{i \phi_{12}} & \ldots & a_{1 N} e^{i \phi_{1 N}} \\
a_{21} e^{i \phi_{21}} & 1 & \ldots & a_{2 N} e^{i \phi_{2 N}} \\
\vdots & \vdots & \ddots & \vdots \\
a_{N 1} e^{i \phi_{N 1}} & a_{N 2} e^{i \phi_{N 2}} & \ldots & 1
\end{array}\right),
\end{eqnarray}
which is used to simultaneously calibrate the XY crosstalks for $N$ qubits \cite{Xu2020}. 

\begin{figure}[htbp]
  \centering
  \includegraphics[width=16cm]{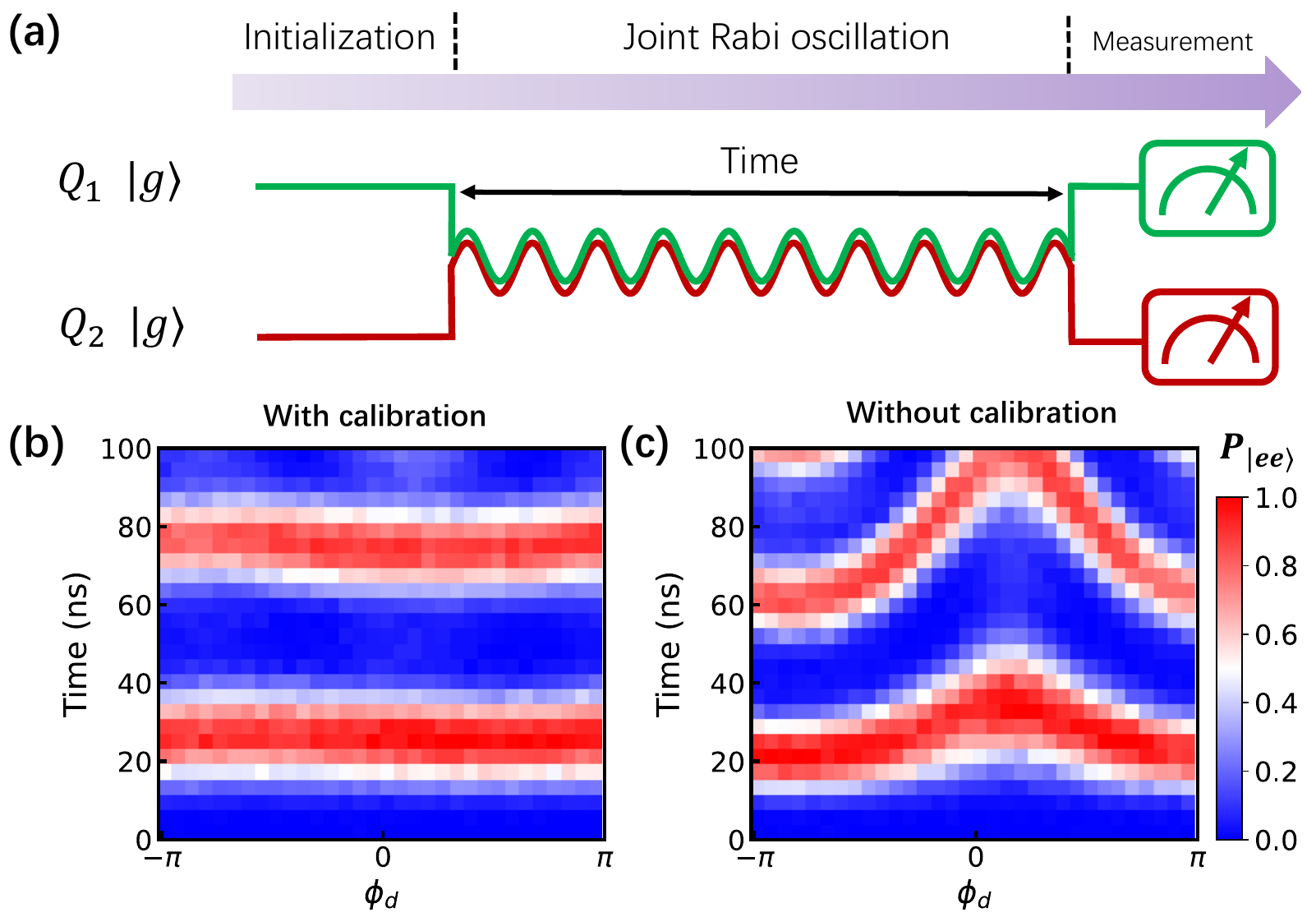} 
  \caption{\textbf{Test of the calibration of the XY crosstalk.} (a) Pulse sequence.
  For testing the calibration of the XY crosstalk between $Q_1$ and $Q_2$, the qubits $Q_1$ and $Q_2$ are initially prepared in the ground state $|gg\rangle$. 
  Then, $Q_1$ and $Q_2$ are biased to the operating point $\omega_o/(2\pi)=5.6895$ MHz by the z-line square waves with the specific ZPAs and we simultaneously input two xy-line microwaves, $A, B=2\pi \times$10 MHz [both with frequency $\omega_o/(2\pi)$] to $Q_1$ and $Q_2$, respectively.
  This process is called the joint Rabi oscillation for short.
  The probability that $Q_1$ and $Q_2$ are simultaneously excited, $P_{|ee\rangle}$, will be measured to determine whether the xy-line microwaves of $Q_1$ and $Q_2$ interfere with each other. 
  (b) (With calibration) Double-excitation population $P_{|ee\rangle}$ versus microwaves phase difference $\phi_d$ and joint Rabi oscillation time. 
  After calibration, it can be seen that the signal of the joint Rabi oscillation is basically not correlated with the $\phi_d$-axis. 
  This is the result we want for $Q_1$ and $Q_2$ where the xy-line microwaves do not interfere.
  (c) (Without calibration) Double-excitation population $P_{|ee\rangle}$ versus microwaves phase difference $\phi_d$ and joint Rabi oscillation time. 
  Here we can see frequency enhancement and attenuation of the joint-Rabi-oscillation signal with respect to the $\phi_d$-axis, which indicates that the interference between the xy-line microwaves still exists.
  }
  \label{test_XYcross}
\end{figure}

To test the effect of this calibration, we prepare $Q_1$ and $Q_2$ both in the ground state $|gg\rangle$ and subsequently bias both qubits to the operating point $\omega_o$ and input the calibrated xy-line microwaves [see Fig. \ref{test_XYcross}(a)]. 
Intuitively, since the two amplitudes of the xy-line microwaves and the initial state of the two qubits are the same, they are dynamically symmetric. 
The probability that they are simultaneously excited, $P_{|ee\rangle}$, should not change with the microwave phase difference $\phi_d$. 
The measured results are shown in Fig. \ref{test_XYcross}(b), where the bright region of excitation can be seen parallel to the $\phi_d$-axis, which is barely independent of the microwave phase difference $\phi_d$, which means that the XY crosstalk has been calibrated.
On the contrary, when the input xy-line microwaves are not calibrated, the measured results $P_{|ee\rangle}$ exhibit non-uniform properties concerning the $\phi_d$-axis in Fig. \ref{test_XYcross}(c).

\subsection{Calibration of the random phase of the initial state} \label{Sec_cal_ini_phase}

\begin{figure}[htbp]
  \centering
  \includegraphics[width=16cm]{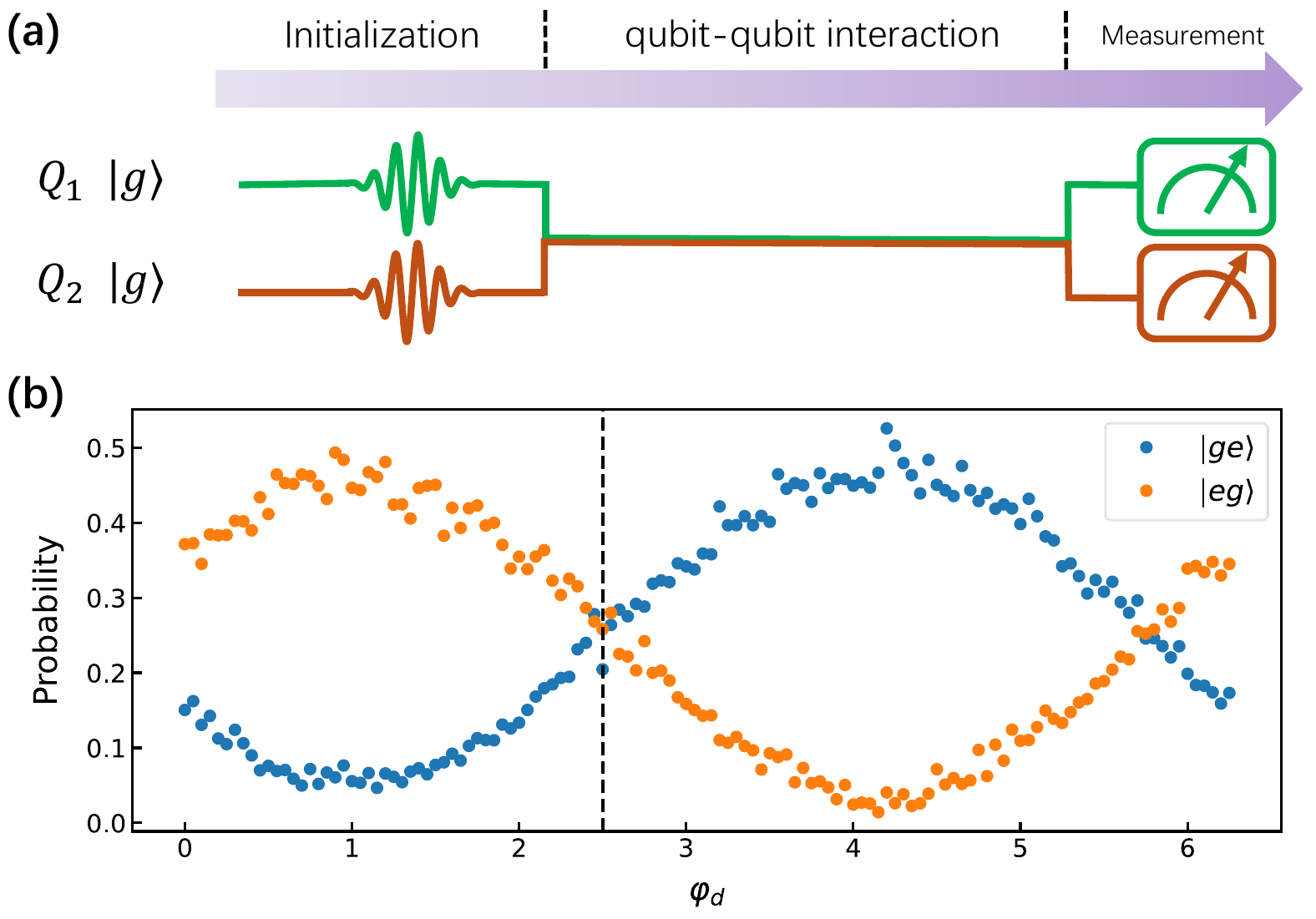} 
  \caption{\textbf{Calibration of the random phase of the initial state.} (a) Pulse sequence. As an example, the qubits $Q_1$ and $Q_2$ are initially prepared in the  state $ (|g\rangle_1+ |e\rangle_1)/\sqrt{2} \otimes (|g\rangle_2+ e^{i\varphi_d}|e\rangle_2)/\sqrt{2}$. 
  Then, we simultaneously bias $Q_1$ and $Q_2$ to the same operating point $\omega_o/(2\pi)=5.6895$ MHz by the z-line square waves with specific ZPAs. 
  The inevitable rising edges of the z-line square waves will turn the initial state into $ (|g\rangle_1+ e^{i\varphi_1} |e\rangle_1)/\sqrt{2} \otimes \left[|g\rangle_2+ e^{i\left(\varphi_2+\varphi_d \right)}|e\rangle_2 \right]/\sqrt{2}$.
  Then these two qubits will interact with each other through second-order interactions.
  At time $t=\pi/(-4\lambda)$, the state will become $\left\{ |gg\rangle+[ie^{i \varphi_1}+e^{i (\varphi_2+\varphi_d ) }]|ge\rangle /\sqrt{2} +[e^{i \varphi_1}+i e^{i (\varphi_2 +\varphi_d)}] |eg \rangle / \sqrt{2} + e^{i (\varphi_1+\varphi_2+\varphi_d ) } |ee\rangle \right\}/2$ and be measured.
  (b) Single-excitation population versus $\varphi_d$. 
  The ideal populations of $|ge\rangle$ and $|eg\rangle$ are $[1\pm \sin (\varphi_d +\varphi_2- \varphi_1)]/4$, respectively.
  We mark $\varphi_d=\varphi_1-\varphi_2$ by a dotted line in the diagram.
  }
  \label{dphasei}
\end{figure}

The initial state of this experiment is prepared at the idle frequency for each qubit. 
After the preparation, each qubit is biased to the operating point $\omega_o/(2\pi)=5.6895$ GHz by its specific z-line square wave. 
This process will accumulate a random phase $\varphi_j$ for each qubit since the square waves have a rising edge. 
So the initial state will be $\otimes_{j=1}^6 (|g\rangle_j+e^{i\varphi_j}|e\rangle_j)/\sqrt{2}$  for the subsequent quasi-adiabatic dynamics, which is unacceptable. 
Therefore, it is necessary to calibrate this random phase $\varphi_j$.
One can lay aside the phase of the first qubit, $\varphi_1$, and then calibrate the initial phase difference between each qubit and the first qubit.  
For example, after the preparation of the initial state $\otimes_{j=1}^2 (|g\rangle_j+e^{i\varphi_j}|e\rangle_j)/\sqrt{2}$, we bias $Q_1$ and $Q_2$ simultaneously to the operating point $\omega_{o}/(2\pi)=5.6895$ GHz. 
The state of these two qubits will be governed by the following Hamiltonian
\begin{eqnarray}
H_{\rm SE}=\lambda |ge\rangle \langle eg | ,
\end{eqnarray}
and becomes $\left[ |gg\rangle+(ie^{i \varphi_1}+e^{i \varphi_2 })|ge\rangle /\sqrt{2} +(e^{i \varphi_1}+i e^{i \varphi_2 }) |eg\rangle / \sqrt{2} + e^{i (\varphi_1+\varphi_2 ) } |ee\rangle \right]/2$ after $t=\pi/(-4\lambda)$, where $\lambda$ is negative. 
When $\varphi_2-\varphi_1 =0$, $|ge\rangle$ and $|eg\rangle$ will have the same probability occupation, 1/4.
In the experiment, we manually change the initial phase difference $\varphi_d$ between $Q_1$ and $Q_2$ to compensate for the random phase difference $\varphi_2-\varphi_1$.
The corresponding pulse sequence is shown in Fig. \ref{dphasei}(a).
The measured population in Fig. \ref{dphasei}(a) can determine the point where $\varphi_d=\varphi_1-\varphi_2=2.5$.
Other random phase differences based on $Q_1$ can analogously be calibrated.
Therefore, the calibrated initial state will thus become $\otimes_{j=1}^6 (|g\rangle_j+e^{i\varphi_1}|e\rangle_j)/\sqrt{2}$.
Later, $\varphi_1$ is scanned (manually changing the initial phase of $Q_1$) to obtain a better GHZ state fidelity.

\subsection{Calibration of the random phase of the continuous microwaves}

\begin{figure}[htbp]
  \centering
  \includegraphics[width=16cm]{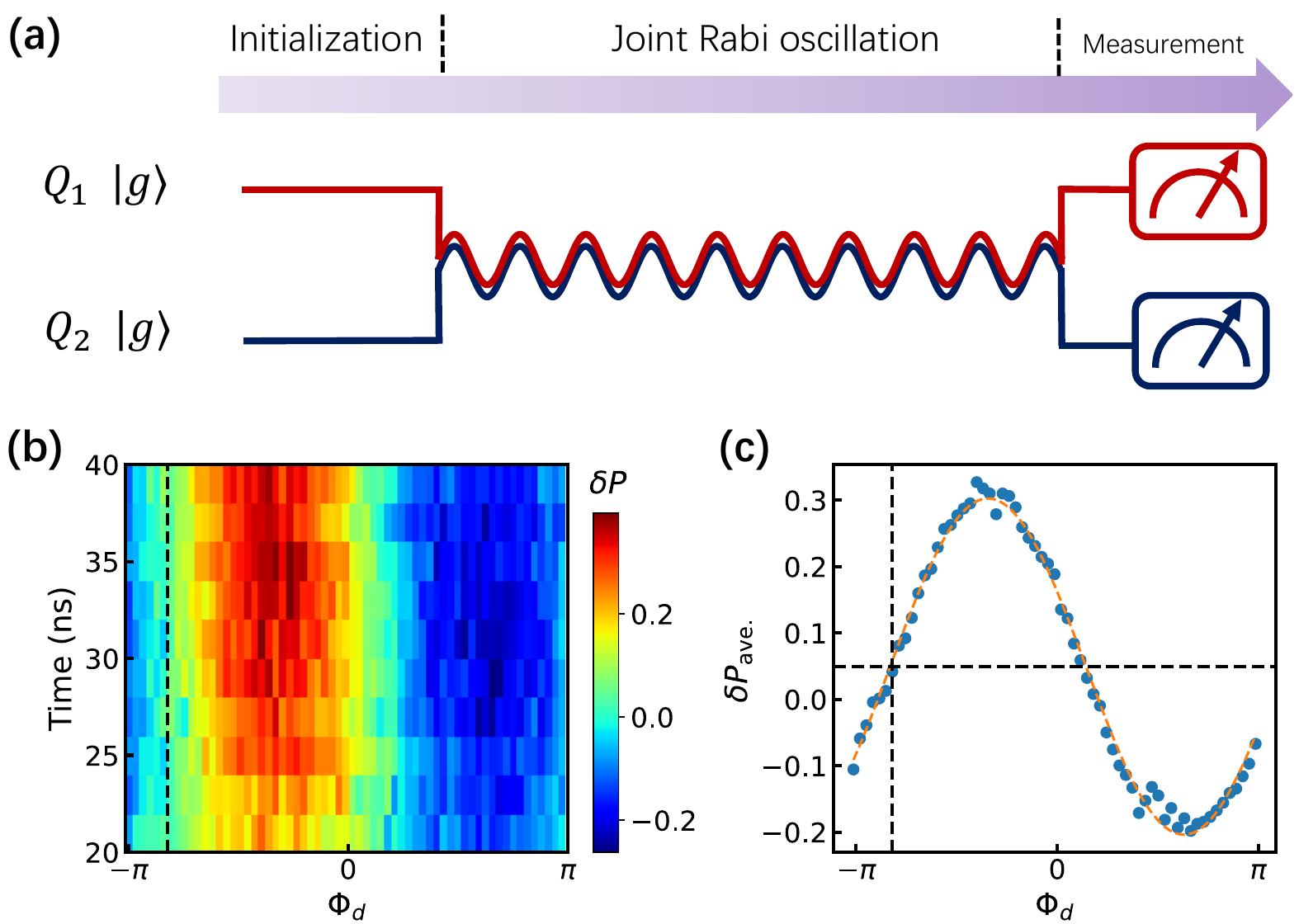} 
  \caption{\textbf{Calibration of the random phase of the xy-line microwaves.} (a) Pulse sequence.
  For example, the qubits $Q_1$ and $Q_2$ are initially prepared in the ground state $|gg\rangle$. 
  Then, $Q_1$ and $Q_2$ are biased to the operating point $\omega_o/(2\pi)=5.6895$ MHz by the z-line square waves with the specific ZPAs and we simultaneously input two xy-line microwaves, $A, B=2\pi \times$5 MHz [both with frequency $\omega_o/(2\pi)$] to $Q_1$ and $Q_2$, respectively.
  The probability difference of the single excitation, $\delta P=P_{|ge\rangle}-P_{|eg\rangle}$, will be measured to determine whether $\Phi_d$ compensates for the real microwave phase difference $\Phi_2-\Phi_1$. 
  (b) The single-excitation-probability difference $\delta P$ versus microwaves phase difference $\Phi_d$ and joint Rabi oscillation time. 
  At around $\Phi_d=-2.55$, marked by a dashed line, the $\delta P$ signal turns from dark to bright. 
  This line means $\Phi_d$ compensates for the real microwave phase difference $\Phi_2-\Phi_1$.
  (c) The averaged single-excitation-probability difference $\delta P_{\rm ave.}$ versus microwaves phase difference $\Phi_d$ and joint Rabi oscillation time. 
  To see more clearly, the signals are averaged within 20-40 ns to obtain $\delta P_{\rm ave.}$. 
  This shows an evident sinusoidal signal.
  A horizontal line can be drawn along the center of the curve, yielding two points on the curve, where the point with a slope $>0$ is the compensation point $\Phi_d=\Phi_1-\Phi_2=-2.55$.
  }
  \label{dOphase}
\end{figure}

Since the lengths of the qubits' xy lines are different (see Fig. 1a in the main text), the xy-line microwave phase $\Phi_j$ for each qubit is different in the experiment.
It is worth mentioning that this phase $\Phi_j$ does not affect the calibration of the XY crosstalk in Sec. \ref{Sec_f2zpa}, because the Rabi signal's oscillation frequency is independent of the microwaves phase.
However, this random phase $\Phi_j$ is not acceptable for the quasi-adiabatic dynamics since one should ensure that the eigenstate of the initial Hamiltonian $H(\tau=0)$ is $|X\rangle$.

The idea of calibrating the random phase of the microwaves is similar to that of calibrating the initial state phase in Sec. \ref{Sec_cal_ini_phase}, which is to calibrate the random phase difference of the microwaves of other qubits versus $Q_1$ first, and then to calibrate the random phase of the microwave of $Q_1$ separately.
The detailed pulse sequence for the calibration is similar to that in Sec. \ref{Sec_Zcross}.
For example, we prepare $Q_1$ and $Q_2$ both in the ground state $|gg\rangle$ and then bias them to the operating point $\omega_o$ and input the calibrated xy-line microwaves with phase difference $\Phi_d$ [see Fig. \ref{dOphase}(a)].
The measured single excitation population difference, $\delta P=P_{|ge\rangle}-P_{|eg\rangle}$ can reflect the microwave phase difference $\Phi_d$.
It can be proved numerically that when $\Phi_d$ precisely compensates for the real microwave phase difference $\Phi_2-\Phi_1$, i.e., $\Phi_d=\Phi_1-\Phi_2$, the $\delta P$ signal in a certain time range will turn from dark to bright, which is shown in Figs. \ref{dOphase}(b) and \ref{dOphase}(c).
These two figures can be utilized to find $\Phi_d=-2.55$.
The random phase difference of the microwaves of other qubits versus $Q_1$ can also be compensated in this way.

\subsection{Calibration of the random phase of the rotation pulses}

\begin{figure}[htbp]
  \centering
  \includegraphics[width=16cm]{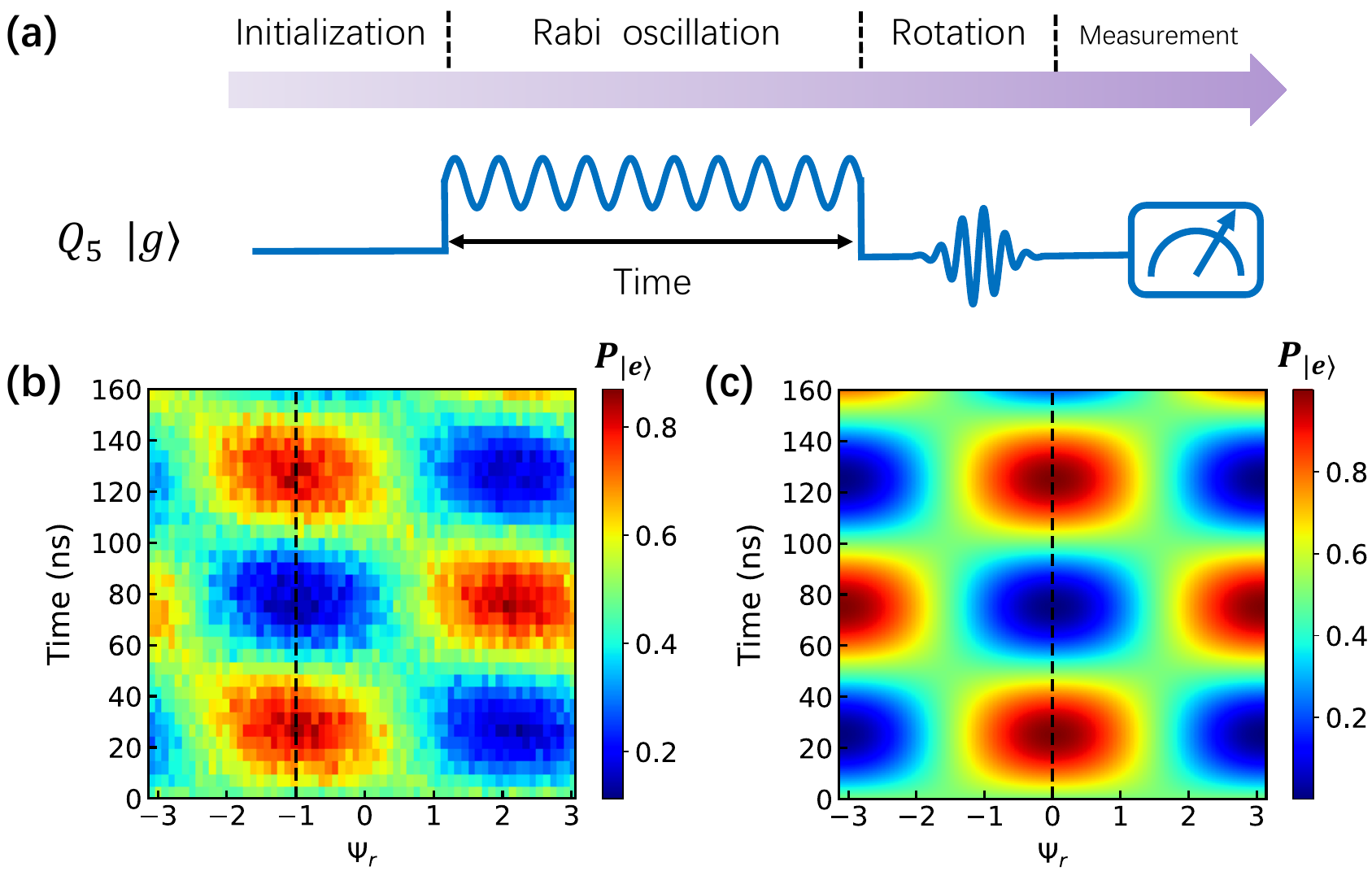} 
  \caption{\textbf{Calibration of the random phase of the rotation pulses.} (a) Pulse sequence.
  Taking $Q_5$ as an example, the qubit $Q_5$ is initially prepared in the ground state $|g\rangle$. 
  Then, we simultaneously drive $Q_5$ with the xy-line microwave and z-line square wave.
  A subsequent $X/2$ pulse with phase $\Psi_r$ is applied before $Q_5$ is measured.
  If $\Psi_r$ compensates for the random phase of the rotation pulse, $\Phi_5$, the measured Rabi oscillation signal will increase from 0.5 and vibrate as a sinusoidal function.
  (b) Measured excited population $P_{|e\rangle}$ versus $\psi_r$ and Rabi oscillation time.
   By comparing with the numerical simulation, one can deduce $\Psi_r=-1.0$.
  (c) Numerical excited population $P_{|e\rangle}$ versus $\psi_r$ and Rabi oscillation time.
  The effect of the falling edge of the z-line square wave is not considered in the numerical simulation, so the compensation phase $\Psi_r=0$, marked by the dashed line.
  }
  \label{rot_phase}
\end{figure}

After the quasi-adiabatic dynamics, we will apply $X/2$ pulses to perform the Wigner tomography or parity oscillation.
However, the z-line square waves in the quasi-adiabatic dynamics usually own a falling edge, resulting in random phases $\Psi_j$ of these $X/2$ pulses.
To calibrate this random phase of the rotation pulse, we can individually simulate a qubit after the quasi-adiabatic dynamics by applying the same z-line pulse ZPA as that in the quasi-adiabatic dynamics and adding an xy-line microwave with frequency $\omega_o/(2\pi)$ and amplitude $A=2\pi\times 5$ MHz. 
This will cause the qubit to proceed Rabi oscillation at frequency $\omega_o/(2\pi)$.
Then, a $X/2$ pulse with phase $\Psi_r$ is added and the qubit will be measured.
These pulse sequences are shown in Fig. \ref{rot_phase}(a).
If $\Psi_r$ compensates for the random phase of the rotation pulse, $\psi_j$, the measured Rabi oscillation signal will increase from 0.5 and vibrate as a sinusoidal function [see Fig. \ref{rot_phase}(c) at $\Psi_d=0$].
The experimental data is shown in Fig. \ref{rot_phase}(b), where one can find that the compensation point is $\Psi_d=-1.0$.
The calibration of the random phase of the rotation pulses for other qubits takes the similar process discussed above.

\clearpage

\section{Effect of nonadiabaticity}

The adiabatic condition is a crucial requirement for ensuring the reliable preparation of the target state during the quasi-adiabatic process. In this section, we explore the effect of nonadiabaticity by performing a detailed numerical simulation. Specifically, we calculate the fidelity of the state evolved under the time-dependent LMG Hamiltonian $H$ with respect to the ideal GHZ state with system size $N=6$. The numerical results confirm that the adiabatic condition is well satisfied during the quasi-adiabatic process, even for a relatively short evolution time of ranging from 149 to 200 ns.

\begin{figure}[h]
    \centering
    \includegraphics[width=16cm]{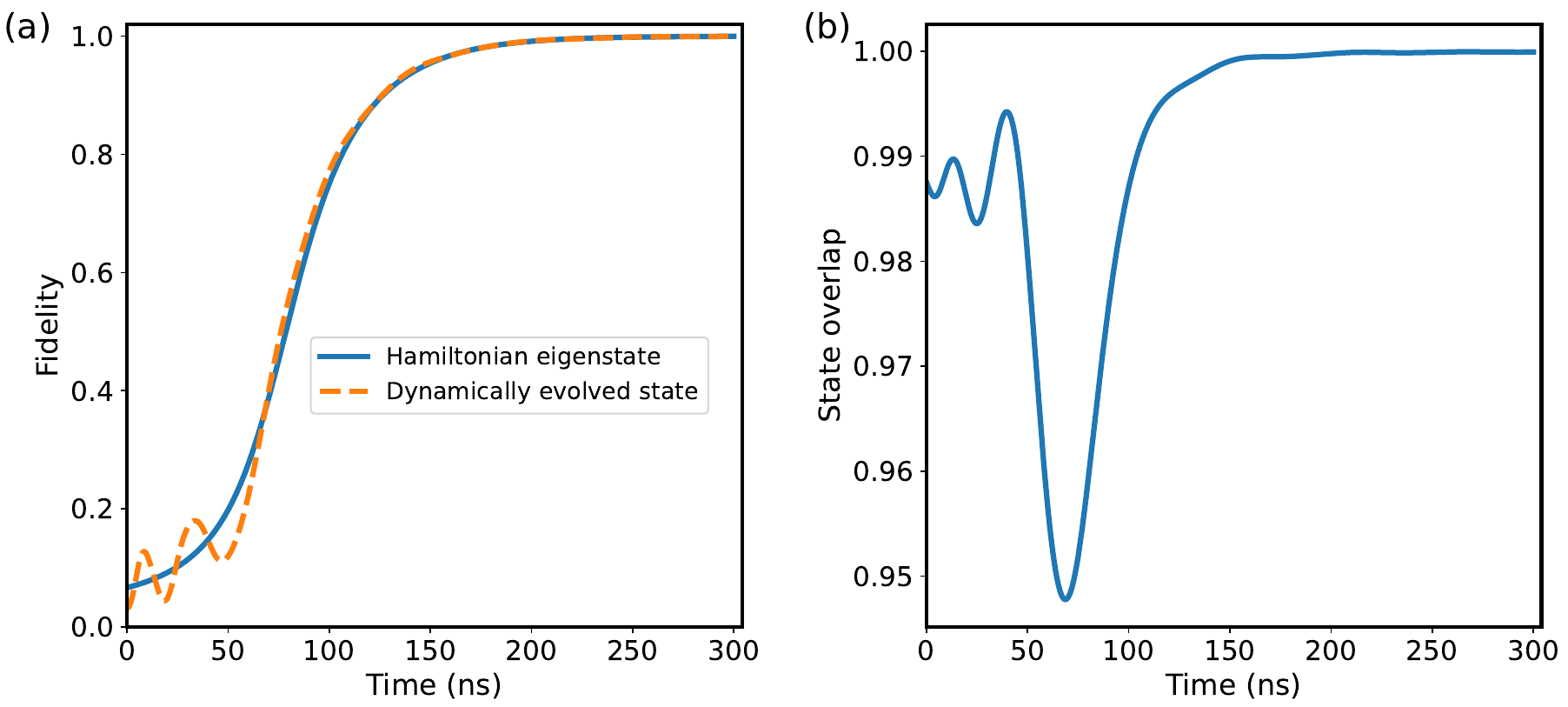}
    \caption{\textbf{Numerical simulation of the adiabatic condition during the quasi-adiabatic process.} (a) Fidelity of the evolved state $|\psi(t)\rangle_{\text{adiabatic}}$ (blue solid line) and the instantaneous eigenstate $|\psi(t)\rangle_{\text{static}}$ (orange dashed line) with respect to the ideal GHZ state. The good agreement between the two curves confirms that the adiabatic condition is well satisfied. (b) Overlap fidelity between $|\psi(t)\rangle_{\text{adiabatic}}$ and $|\psi(t)\rangle_{\text{static}}$, showing values close to unity throughout the quasi-adiabatic process. The results validate the adequacy of the evolution time (149-200 ns) for achieving adiabaticity.}
    \label{fig:nonadiabaticity}
\end{figure}

We first consider the fidelity ${\cal F}$ between the evolved state $|\psi(t)\rangle_{\text{adiabatic}}$ and the ideal GHZ state, which is defined as
\begin{equation}
{\cal F}=\left|\langle\text{GHZ}|\psi(t)\rangle_{\text{adiabatic}}\right|^2.
\end{equation}
The blue solid line in Fig.~\ref{fig:nonadiabaticity}(a) shows the time evolution of ${\cal F}$ under the time-dependent Schr\"odinger equation with $H$ in Eq. (\ref{eq:LMG}) with $\Omega=\Omega^0 \exp(-t/t_f)$, where $\Omega^0/(2\pi)=40$ MHz and $t_f=60$ ns.
To further verify the adiabatic nature of the process, we compute the fidelity of the instantaneous eigenstate $|\psi(t)\rangle_{\text{static}}$ of $H$ with respect to the ideal GHZ state. The result, shown as the orange dashed line in Fig.~\ref{fig:nonadiabaticity}(a), shows good consistency with the blue solid line. This agreement strongly suggests that the evolved state closely follows the ideal  eigenstate of the Hamiltonian throughout the quasi-adiabatic process.

To gain deeper insights into the dynamics, we also calculate the overlap between $|\psi(t)\rangle_{\text{adiabatic}}$ and $|\psi(t)\rangle_{\text{static}}$, using the fidelity definition:
\begin{equation}
    {\cal F}_{\text{overlap}} = \left|\langle\psi(t)|_{\text{static}} |\psi(t)\rangle_{\text{adiabatic}}\right|^2.
\end{equation}

As shown in Fig.~\ref{fig:nonadiabaticity}(b), the overlap fidelity ${\cal F}_{\text{overlap}}$ remains very close to unity throughout the evolution, further confirming the validity of the adiabatic condition. Notably, we observe small oscillations ($<0.05$) in ${\cal F}_{\text{overlap}}$ near the critical regime of the quasi-adiabatic process, which result from the finite evolution time and the associated nonzero rate of change of the Hamiltonian parameters. In the numerical simulations, increasing both the evolution time $t$ and the characteristic time $t_f$ can suppress these oscillations, bringing ${\cal F}_{\text{overlap}}$ closer to unity. However, under the current experimental parameters, an evolution time of approximately 149–200 ns is sufficient to achieve adiabaticity while balancing the effects of decoherence. Extending the evolution time may reduce nonadiabatic effects but would introduce additional challenges due to these decoherence mechanisms. Therefore, the chosen parameters provide a practical compromise between achieving adiabaticity and ensuring experimental feasibility.

These results demonstrate that the adiabatic condition is well satisfied under the chosen parameters. Therefore, the quasi-adiabatic process employed in our experiment can ensure adiabatic preparation of the target GHZ state.

\clearpage

\section{Simulation of the System Dynamics}

\begin{figure}[h]
    \centering
    \includegraphics[width=0.97\linewidth]{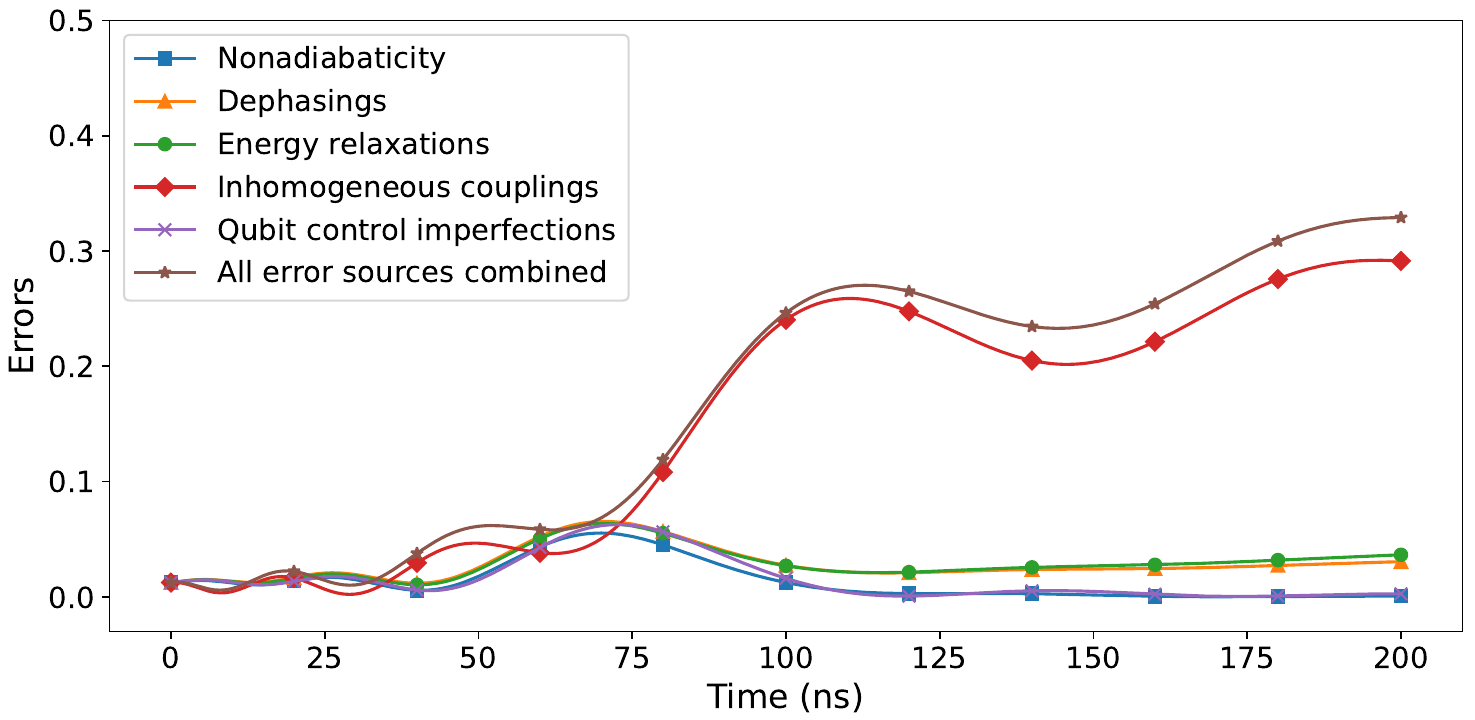}
    \caption{\textbf{Numerical simulation of error evolution under different experimental imperfections.} 
    The error evolutions under various error sources are as follows: nonadiabatic effects, dephasings, energy relaxations, inhomogeneous coupling, qubit control imperfections, and the combined effects of all error sources. 
    Nonadiabatic effects and qubit control imperfections have minor contributions to the errors, which remain small throughout the evolution. Decoherence mechanisms (dephasings and energy relaxations) cause moderate increases in errors. Inhomogeneity of the intraqubit couplings is the dominant error source, which contributes about 85\% of the total error. }
    \label{fig:errors}
\end{figure}

To quantify the errors caused by different experimental imperfections, we perform numerical simulations to analyze the impacts of the nonadiabatic effect, qubits' dephasings and energy relaxations, inhomogeneity of the intraqubit couplings, and imperfections in qubit control. The results are presented in Fig.~\ref{fig:errors} and Table~\ref{tab:errors}.

Each simulation considers only a single error source to isolate its contributions to the error evolution with respect to the instantaneous Hamiltonian eigenstate $|\psi(t)\rangle_{\mathrm{static}}$. The error is defined as the infidelity between the evolved state $\rho(t)$ (density matrix) and the eigenstate $|\psi(t)\rangle_{\mathrm{static}}$ at time $t$:
\[
\text{Error} = 1 - \rm{Tr}\left[  \rho(t) | \psi(t)\rangle_{\mathrm{static}}\langle \psi(t)|_{\mathrm{static}}\right].
\]
This approach ensures a clear understanding of how each type of imperfection affects the system dynamics.

\textbf{Nonadiabaticity:} As shown in Fig.~\ref{fig:errors}, the system well follows the eigenstate of the control Hamiltonian during the quasi-adiabatic process if other error sources are not taken into account. The error due to nonadiabaticity is only 0.002 at 149 ns, which is negligible compared with other errors.

\textbf{Dephasings:} The error caused by dephasings is estimated with dephasing times ($T_2^{\rm c}$) with the continuous dynamical decoupling protection \cite{Guo2018}, measured under a continuous drive of $2\pi\times5.4\,\mathrm{MHz}$  (corresponding to the average continuous driving strength in the first 149 ns of the experiment). The measured values of $T_2^{\rm c}$ for the six qubits are 8.8, 12.8, 9.6, 9.3, 11.3, and 24.4 $\mu$s, respectively. As shown in Fig.~\ref{fig:errors}, due to the inherent dynamical decoupling enabled by continuous driving, dephasing noises only cause a slight error, which is equal to 0.024 at 149 ns.

\textbf{Energy relaxations:} The error caused by energy relaxations of the qubits is simulated with the values of $T_1$ listed in Table \ref{paras}. The result is 0.027 at 149 ns.

\textbf{Inhomogeneous couplings:} Due to imperfection of the fabrication of the superconducting circuit, the qubits are not homogeneous coupled to one another. This inhomogeneity arises from the asymmetry of couplings with the bus resonator, as shown in Table \ref{paras}, and from the asymmetry of direct qubit-qubit couplings. The effective qubit-qubit couplings listed in Table \ref{tab:coupling}, each of which is measured by tuning the two corresponding qubits to the operation frequency and observing their population evolution. As shown in Fig.~\ref{fig:errors}, the error caused by this inhomogeneity is much larger than other errors, reaching a value of 0.203 at 149 ns.

\textbf{Qubit control imperfection:} The quasi-adiabatic process is realized by varying the strength of the drive applied to each qubit as $\Omega(t)=\Omega^0\exp(-t/t_{f})$, where $\Omega^0$ is set to be $2\pi \times$ 40 MHz. Due to the control imperfection, there is a deviation between the real driving strength ($\Omega_r^0$) and the preset value. For $\Omega_r^0=0.99 \Omega^0$, the resulting error is 0.005 at 149 ns.

We also perform a simulation with the master equation, where all the aforementioned imperfections are included. The resulting fidelity is 0.774 at 149 ns.
It should be noted that the ideal Hamiltonian eigenstate at 149 ns is not exactly a GHZ state, with an overlap of 0.959. According to the simulation, the state at 149 ns has a fidelity of 0.774 with respect to the GHZ state, which well agrees with the measured result, 0.770.

\begin{table}[h]
    \centering
    \begin{tabular}{|@{\hspace{0.5cm}}c@{\hspace{0.5cm}}|@{\hspace{0.5cm}}c@{\hspace{0.5cm}}|@{\hspace{0.5cm}}c@{\hspace{0.5cm}}|@{\hspace{0.5cm}}c@{\hspace{0.5cm}}|@{\hspace{0.5cm}}c@{\hspace{0.5cm}}|@{\hspace{0.5cm}}c@{\hspace{0.5cm}}|@{\hspace{0.5cm}}c@{\hspace{0.5cm}}|} 
        \hline
              & $Q_1$  & $Q_2$ & $Q_3$ & $Q_4$ & $Q_5$ & $Q_6$ \\
        \hline
        $Q_1$ &       &         &        &        &       &     \\
        \hline
        $Q_2$ & 3.28  &         &        &        &       &     \\
        \hline
        $Q_3$ & 3.06  &  2.98   &        &        &       &     \\
        \hline
        $Q_4$ & 3.58  &  3.50   &  2.50  &        &       &     \\
        \hline
        $Q_5$ & 1.97  &  3.47   &  3.17  &  3.68  &       &     \\
        \hline
        $Q_6$ & 2.46  &  2.43   &  1.65  &  2.12  &  2.53 &   \quad  \quad  \  \\
        \hline
    \end{tabular}
    \caption{\textbf{Measured effective qubit-qubit coupling strengths (in units of $\bm {2\pi}$ MHz)}. These values reflect the inhomogeneity caused by asymmetry of couplings with the bus resonator and direct qubit-qubit interactions. The coupling strength between each pair of qubits is measured by tuning the two qubits to their operation frequencies and analyzing their population evolution. }
    \label{tab:coupling}
\end{table}

\begin{table}[h]
    \centering
    \begin{tabular}{c@{\hspace{1cm}}c} 
        \hline
        \textbf{Error Sources} & \textbf{Errors}  \\
        \hline
        Nonadiabaticity & $0.002$  \\
        Dephasings & $0.024$  \\
        Energy relaxations & $0.027$  \\
        Inhomogeneous couplings & $0.203$  \\
        Qubit control imperfections & $0.005$  \\
        All errors combined & $0.235$ \\
        \hline
    \end{tabular}
    \caption{\textbf{Summary of errors under different experimental imperfections.} Errors are calculated as the infidelity between the simulated state and the instantaneous Hamiltonian eigenstate at $149$ ns.  Each simulation considers only one error source, except for the all errors combined case, which includes the cumulative effects of all imperfections.}
    \label{tab:errors}
\end{table}

\clearpage

\section{Evolution of the multiqubit quantum correlations for different system sizes}

To investigate the dependence of quantum-mechanical features on the system size, we perform numerical simulations of the transverse quantum correlation evolution among qubits for various qubit numbers \(N\).  In the followed simulations, the fidelity is used as a measure to quantify the multiqubit quantum correlations. By evaluating the fidelity, we can assess how the coherence among qubits evolves under different system sizes and the effects of decoherence.
The simulations are conducted in the symmetrized subspace, which significantly reduces the computational complexity. Decoherence, including qubit energy relaxations and dephasings, are incorporated into the master equation within the symmetrized framework. The master equation governing the system dynamics is given by: 
\begin{equation}
\frac{d\rho}{dt} = -i[H, \rho] + \sum_{i=1}^N \frac{1}{T_1} \mathcal{D}[S_-]\rho + \sum_{i=1}^N \frac{2}{T_\phi} \mathcal{D}[S_z/ \sqrt{N}]\rho,
\end{equation}
where \(H\) denotes the LMG Hamiltonian in Eq. (\ref{eq:LMG}), and \(\mathcal{D}[{\cal A}]\rho = {\cal A}\rho {\cal A}^\dagger - \frac{1}{2}\{{\cal A}^\dagger {\cal A}, \rho\}\) represents the Lindblad superoperator describing dissipative dynamics characterized by the energy relaxation time \(T_1\) and the pure dephasing time \(T_\phi\).
The parameter \(T_\phi\) is related to \(T_1\) and the continuous-driving dephasing time \(T_2^{\rm c}\) through the relation:
\begin{align}
\frac{1}{T_2^{\rm c}} = \frac{1}{2T_1} + \frac{1}{T_\phi}.
\end{align}
The decoherence parameters (\(T_1\) and \(T_2^{\rm c}\)) and intraqubit couplings used in the simulation correspond to the average values of our experimental system.
Note the transverse drivings, $\Omega(t) = \Omega^0 e^{-t/t_f}$ scale linearly with $N$.

\begin{figure}[h]
    \centering
    \includegraphics[width=16cm]{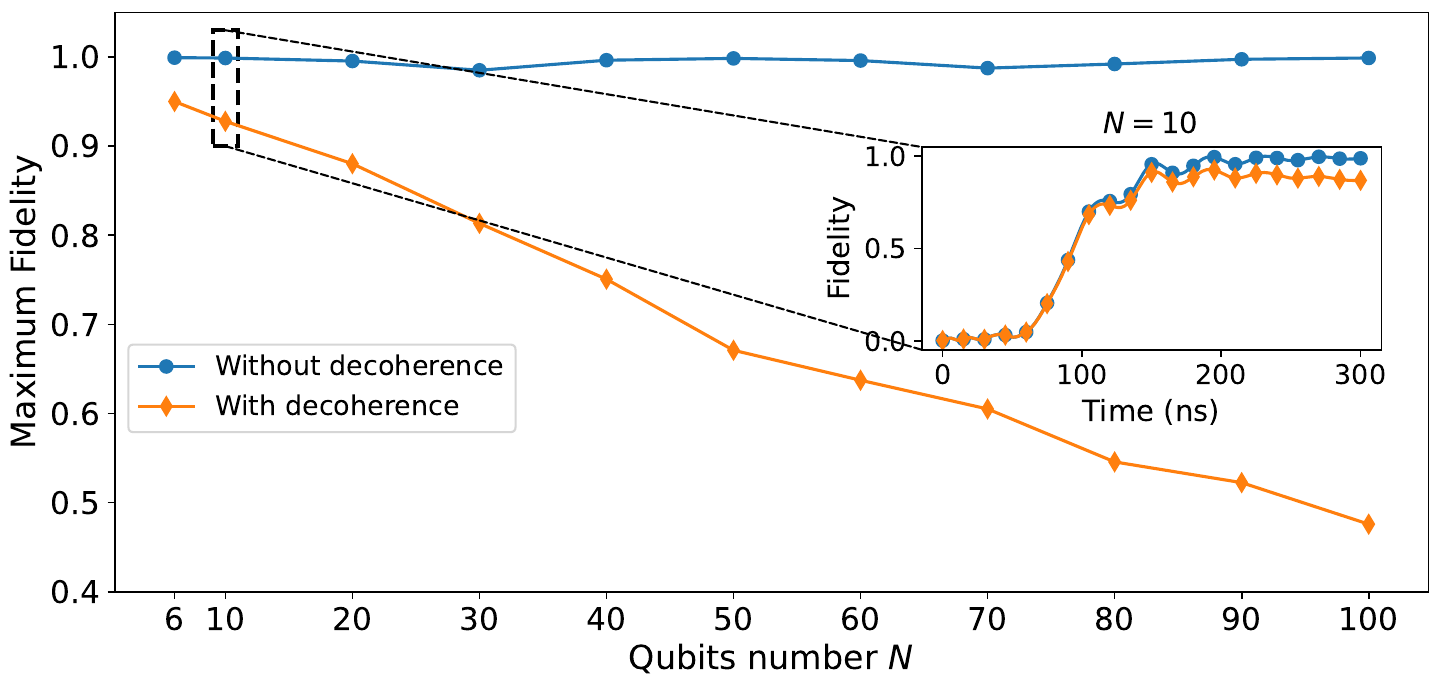}
    \caption{\textbf{Maximum fidelity versus the qubit number \( \bm N\), simulated in the symmetrized subspace.} The diamond-marked and circle-marked lines correspond to the results for qubits with decoherence (including energy relaxations and dephasings) and without decoherence, respectively. The inset shows the time evolution of fidelity for \(N=10\), where the effects of decoherence are explicitly compared. The decoherence effects are incorporated using the Lindblad master equation in the symmetrized space. The results verify that the quantum-mechanical feature of SSB diminishes as \(N\) increases, consistent with the linear scaling of the decoherence rate.
    }
    \label{Nup_Fdown}
\end{figure}

Figure \ref{Nup_Fdown} shows the maximum fidelity as a function of the qubit number \(N\), alongside an inset illustrating the time evolution of fidelity for \(N=10\). The fidelity decreases monotonically with increasing \(N\), reflecting the linear scaling of the decoherence rate for GHZ states. This demonstrates that the quantum-mechanical feature of SSB vanishes in the thermodynamic limit as \(N \to \infty\).

The numerical results confirm that for finite-size systems, the quantum coherence between symmetry-breaking state components persists during the quantum phase transition. However, as \(N\) increases, the decoherence effects dominate, and the quantum-mechanical features gradually diminish. These simulations emphasize the fundamental distinction between finite-size quantum many-body systems and their infinite-size counterparts.

\clearpage

\bibliography{references}